\documentclass[english]{article}
\usepackage[latin1]{inputenc}
\usepackage{amsmath}
\usepackage{amssymb}

\makeatletter

\usepackage[abbr,dcucite]{harvard}
\usepackage{graphicx}
\usepackage{babel}

\usepackage{html}  

\renewcommand*{\sectionmark}[1]{} 
\renewcommand*{\subsectionmark}[1]{} 

\usepackage{babel}
\makeatother
\begin{document}
\renewcommand{\vec}[1]{\mathbf{#1}}
\newcommand{\ket}[1]{|#1\rangle}

\newcommand{\bra}[1]{\langle#1|}

\newcommand{\braopket}[3]{\left\langle #1\left|#2\right|#3\right\rangle }

\newcommand{\braket}[2]{\langle#1|#2\rangle}

\newcommand{\bsigma}{\boldsymbol{\sigma}}

\newcommand{\brho}{\boldsymbol{\rho}}

\newcommand{\bzeta}{\boldsymbol{\zeta}}

\newcommand{\bvarepsilon}{\boldsymbol{\varepsilon}}

\newcommand{\bkappa}{\boldsymbol{\kappa}}

\newcommand{\btau}{\boldsymbol{\tau}}

\newcommand{\bpi}{\boldsymbol{\pi}}

\newcommand{\dotvec}[1]{\vec{\dot{#1}}}

\newcommand{\ShermanBmAvg}{\gamma}

\newcommand{\cSOLeff}{c^{*}}

\newcommand{\fExpCoeffZero}{C_{k}}

\newcommand{\fExpCoeffLinearSO}{D_{k}}

\newcommand{\SOcoupling}{\lambda}

\newcommand{\SOcouplingVac}{\SOcoupling_{\mathrm{vac}}}

\newcommand{\SOcouplingEl}{\SOcoupling}

\newcommand{\SOcouplingHole}{\SOcoupling_{\mathrm{h}}}

\newcommand{\SOcouplingVal}{\lambda_{\mathrm{v}}}

\newcommand{\polAHE}{\vec{P}_{\mathrm{AH}}}

\newcommand{\JAHE}{\vec{J}_{\mathrm{AH}}}

\newcommand{\jCCC}[3]{j_{#1,\,#3}^{#2}}

\newcommand{\jSH}{\vec{j}}

\newcommand{\jSHc}{j}

\newcommand{\jSHSJ}{\jSH_{\,\mathrm{SJ}}}

\newcommand{\spinCurrentVec}[1]{\vec{j}^{\,#1}}

\newcommand{\spinCurrentComponent}[2]{j_{#2}^{#1}}

\newcommand{\spinCurrentOperator}[2]{\hat{j}_{#2}^{#1}}

\newcommand{\spinHallConductivity}{\sigma^{\mathrm{SH}}}

\newcommand{\sigmaSHSS}{\sigma_{\mathrm{SS}}^{\mathrm{SH}}}

\newcommand{\unitVector}[1]{\hat{\vec{#1}}}

\newcommand{\crossSecSpin}{d\negmedspace\!\stackrel{\leftrightarrow}{\sigma}\negmedspace\!}

\newcommand{\Egap}{E_{\mathrm{0}}}

\newcommand{\DeltaSO}{\Delta_{0}}

\newcommand{\Heff}{H_{\mathrm{eff}}}

\newcommand{\HSOlbl}[2]{H_{#1,\,#2}}

\newcommand{\Hintrinsic}{H_{\mathrm{int}}}

\newcommand{\HDresselhausThreeD}{\HSOlbl{\mathrm{D}}{3\mathrm{d}}}

\newcommand{\HDresselhausk}{H_{\beta}}

\newcommand{\HRashbaEl}{H_{\alpha}}

\newcommand{\HDresselhausTwoDkkk}{\HSOlbl{\mathrm{D}}{2\mathrm{d}}}

\newcommand{\HRashbaHTwoD}{H_{\alpha,\mathrm{h}}}

\newcommand{\Hextrinsic}{H_{\mathrm{ext}}}

\newcommand{\HextrinsicEl}{\HSOlbl{\mathrm{ext}}{\mathrm{e}}}

\newcommand{\HextrinsicHole}{\HSOlbl{\mathrm{ext}}{\mathrm{h}}}

\newcommand{\HextrinsicVal}{\HSOlbl{\mathrm{ext}}{\mathrm{v}}}

\newcommand{\HstrainEl}{\HSOlbl{\varepsilon}{\mathrm{e}}}

\newcommand{\HstrainVal}{\HSOlbl{\varepsilon}{\mathrm{v}}}

\newcommand{\HstrainHole}{\HSOlbl{\varepsilon}{\mathrm{h}}}

\newcommand{\couplingDresselhausThreeD}{\mathcal{B}}

\newcommand{\Angstrom}{\mbox{\AA}}

\global\emergencystretch = 0.2\hsize

\title{Theory of Spin Hall Effects in Semiconductors }

\author{Hans-Andreas Engel, Emmanuel I. Rashba, and Bertrand I. Halperin }

\maketitle
Department of Physics, Harvard University, Cambridge, Massachusetts
02138 

\begin{abstract}
Spin Hall effects are a collection of phenomena, resulting from spin-orbit
coupling, in which an electrical current flowing through a sample
can lead to spin transport in a perpendicular direction and spin accumulation
at lateral boundaries. These effects, which do not require an applied
magnetic field, can originate in a variety of intrinsic and extrinsic
spin-orbit coupling mechanisms and depend on geometry, dimension,
impurity scattering, and carrier density of the system---making the
analysis of these effects a diverse field of research. In this article,
we give an overview of the theoretical background of the spin Hall
effects and summarize some of the most important results. First, we
explain effective spin-orbit Hamiltonians, how they arise from band
structure, and how they can be understood from symmetry considerations;
including intrinsic coupling due to bulk inversion or structure asymmetry
or due to strain, and extrinsic coupling due to impurities. This leads
to different mechanisms of spin transport: spin precession, skew scattering,
and side jump. Then we present the kinetic (Boltzmann) equations,
which describe the spin-dependent distribution function of charge
carriers, and the diffusion equation for spin polarization density.
Next, we define the notion of spin currents and discuss their relation
to spin polarization. Finally, we explain the electrically induced
spin effects; namely, spin polarization and currents in bulk and near
boundaries (the focus of most current theoretical research efforts),
and spin injection, as well as effects in mesoscopic systems and in
edge states. 
\end{abstract}
\begin{quotation}
\end{quotation}

\section{Introduction}

In the simplest version of a spin Hall effect, an electrical current
passes through a sample with spin-orbit interaction, and induces a
spin polarization near the lateral edges, with opposite polarization
at opposing edges \cite{DPpolarization}. This effect does not require
an external magnetic field or magnetic order in the equilibrium state
before the current is applied. If conductors are connected to the
lateral edges, spin currents can be injected into them. Electrical
current in a sample can also produce a bulk spin polarization, far
from the edges, which is not generally classified \textit{per se}
as a spin Hall effect, though it is intimately related and is an important
ingredient of spin Hall calculations. 

Spin Hall effects have received a great deal of theoretical attention
recently, in part because the subject includes ingredients of spintronics,
electrical generation, transport, and control of nonequilibrium spin
populations, and also because analysis has shown that the problem
has remarkable subtlety. Theoretical efforts were also fueled by recent
experimental observations of these effects.

In the following we consider semiconductors, where the various mechanisms
of spin-orbit coupling are well-known and can be roughly classified
into two categories. First, the extrinsic mechanism is only present
in the vicinity of impurities and leads to spin-dependent scattering,
including Mott skew scattering. Second, the intrinsic spin-orbit coupling
remains finite away from impurities and can be understood as a (often
spatially homogeneous) spin-orbit field inherent in the band structure.
Furthermore, the spin-orbit couplings are strongly symmetry dependent
and are therefore different for electron and for hole carriers, and
are different for two and three dimensional systems. Thus, the microscopic
processes involved in generating a spin Hall effect can depend critically
on such system properties. Initially, there was hope that spin transport
theory can be formulated in terms of universal spin currents that
would simplify our understanding; however, it turned out that there
is no such universality. As there is no unique description of the
spin Hall effect, we should rather refer to it as a set of spin Hall
effects. 

There is already a vast amount of theoretical literature available
on spin Hall effects and on the related spin currents. It is beyond
the scope of this article to provide a historical overview or to give
an explanation of all the theoretical techniques used. We rather provide
an overview of the various mechanisms, explain them using intuitive
and qualitative physical pictures, and give a summary of some key
theoretical descriptions and results. 

In contrast, the number of experiments on spin Hall effects is small
and an overview is straightforward. In the experiment by \citeasnoun{KatoSpinHall},
electrical currents in three-dimensional $n$-GaAs layers (2$\mu$m
thick) induced a spin Hall effect, which was optically detected via
Kerr microscopy. Measurements in strained samples showed little dependence
of the effect on the crystal orientation and it was concluded that
the extrinsic mechanism proposed by \citeasnoun{DPpolarization} was
causing the spin Hall effect \cite{KatoSpinHall}. Indeed, the experimental
data can be described with reasonable accuracy by the extrinsic mechanism
in a model based on scattering by screened Coulomb impurities \cite{EHR_extrinsic},
as well as one based on short-range scatterers \cite{TseDasSarmaExtrinsic}.
(See~Sec\@.~\ref{sub:Bulk-Spin-Currents}.) Similar experiments
in ZnSe \cite{SternZnSe} were also in agreement with theory, and
a spin Hall effect was observed at room temperature. In another experiment,
\citeasnoun{Wunderlich05} observed a spin Hall effect in two-dimensional
layers of $p$-GaAs by detecting the polarization of recombination
radiation at the edges of the sample. They ascribed this effect to
the intrinsic mechanism, which is consistent with the magnitude of
the observation \cite{Schliemann_2DHolesk3,NomuraWunderlich}. Furthermore,
in measurements on a two-dimensional electron system in an AlGaAs
quantum well, a spin Hall effect was also observed and ascribed to
the extrinsic mechanism \cite{SihSHE}. Finally, \citeasnoun{ValenzuelaSH}
observed a reciprocal spin Hall effect in Al, where a spin current
induced a transverse voltage via the extrinsic mechanism \cite{Hirsch99}.

In addition to theoretical works in the traditional sense, there is
also a large amount of numerical simulations on concrete realizations
of disordered systems, e.g., on a finite lattice. It has generally
been difficult to make direct comparisons between numerical simulations
and theoretical predictions, in part because theoretical works usually
assume that the spin-orbit splittings are much less than the Fermi
energy, while simulations tend to employ larger spin orbit splittings
in order to obtain numerically significant results. See for example
\cite{AndoTightB,ShengShengTing,Nikolic05,ShenSimulations}. 

In Section~\ref{sec:Spin-Orbit-SC} below, we review the mechanisms
for spin-orbit coupling in the semiconductors of interest, and we
discuss forms of the effective Hamiltonians that describe the carriers
in various situations. In Section~\ref{sec:SO-Mechanisms}, we see
how the various effective Hamiltonians can influence spin transport
and accumulation. We discuss spin precession, produced by the intrinsic
spin-orbit coupling, as well as skew scattering and the so-called
side-jump effect, resulting from the extrinsic spin-orbit coupling.
We introduce Boltzmann-type kinetic equations which can describe spin-transport
and accumulation in various situations, and we discuss the simpler
spin and charge diffusion equations which can typically be used in
situations where the spin relaxation rate is much slower than momentum
relaxation.

In Section~\ref{sec:Electrically-induced-SpinEffects}, we discuss
explicitly the spin polarization and spin transport arising from an
electrical current in a semiconductor with spin-orbit coupling. We
introduce the notion of a spin current and the spin Hall conductivity,
and we discuss results that have been obtained for these quantities
in various situations. We also discuss a relation between the spin
Hall conductivity and the so-called anomalous Hall effect that can
result from spin-orbit coupling in a ferromagnet or in semiconductor
with a spin polarization induced by an external magnetic field.

Spin currents are not directly observed in experiments, however. If
spin relaxation rates are slow, one may expect that spin-currents
with a non-zero divergence can lead to observable local spin polarizations,
in which relaxation of excess spin balances the accumulation of spin
that is transported into a region by the spin current. Furthermore,
boundary effects may be important and non-trivial; in the presence
of an electric current, spin-polarization may be generated directly
at a sample boundary. These issues are discussed in Subsection~\ref{sub:Spin-at-boundaries}.
We also discuss briefly mesoscopic systems, where all parts of the
sample are close to a boundary.

A different type of spin Hall effect, associated with edge states,
has been predicted to occur in certain systems that are insulating
in the bulk, where the topology of the band structure has been altered
due to spin orbit coupling. In Section~\ref{sec:SHEdgeStates}, we
discuss this concept, along with the possibility that such effects
may occur and be observable in a number of materials.

\section{Spin-orbit coupling in semiconductors}

\label{sec:Spin-Orbit-SC}
\newcommand{\Jabs}{J}

\newcommand{\Jz}{J_{z}}

\newcommand{\JabsHoles}{\Jabs=3/2}

\newcommand{\Vtotal}{\tilde{V}}

\newcommand{\Vperiodic}{V_{\mathrm{cr}}}

\newcommand{\Vslow}{V}

\newcommand{\SOcouplingPseudoSpin}{\SOcoupling}

For a non-relativistic electron in vacuum, the Dirac equation can
be reduced to the Pauli equation, describing a two-component spinor
and containing the Zeeman term. The Pauli equation also contains relativistic
corrections---including the spin-orbit coupling\begin{equation}
H_{\mathrm{SO,\, vac}}=\SOcouplingVac\:\bsigma\cdot\left(\vec{k}\times\nabla\Vtotal\right).\label{eq:HVacuum}\end{equation}
 Here, we used $\SOcouplingVac=-\hbar^{2}/4m_{0}^{2}c^{2}\approx-3.7\times10^{-6}\:\Angstrom{}^{2}$,
vacuum electron mass $m_{0}$, velocity of light $c$, and $\vec{k}=\vec{p}/\hbar$.
In a semiconductor, we split the total potential $\Vtotal=\Vperiodic+\Vslow$
into the periodic crystal potential $\Vperiodic$ and an aperiodic
part $\Vslow$, which contains the potential due to impurities, confinement,
boundaries, and external electrical field. One then tries to eliminate
the crystal potential as much as possible and to describe the charge
carriers in terms of the band structure.  The simplest systems of
this sort can be exemplified by electrons in cubic direct-gap semiconductors.
Then, the minimum of the energy spectrum is usually near the center
of the Brillouin zone, and the two-fold Kramers degeneracy is the
only degeneracy of the spectrum at $k=0$. It follows from symmetry
arguments, that for slow electrons in such crystals, and for slow
carriers (electrons and holes) in high symmetry two-dimensional systems,
the effective single-particle Hamiltonian is 
\newcommand{\energyOfK}[1]{\epsilon_{#1}}
\begin{align}
\Heff & =\energyOfK{k}+\Vslow+\Hintrinsic+\Hextrinsic,\label{eq:Heffective}\\
\Hintrinsic & =-\frac{1}{2}\,\vec{b}\left(\vec{k}\right)\cdot\bsigma,\label{eq:HIntrinsic}\\
\Hextrinsic & =\SOcouplingPseudoSpin\:\bsigma\cdot\left(\vec{k}\times\nabla\Vslow\right),\label{eq:Hextrinsic}\end{align}
 where $\vec{k}$ is the crystal wave vector relative to the zone
center, and we assumed that $\Vslow$ is only slowly varying on the
scale of the lattice constant. Here $\bsigma$ is the vector of Pauli
matrices for the pseudo spin-$\frac{1}{2}$ of the Kramer's doublet
at $k=0$; it is customary called a spin-$\frac{1}{2}$ system. $\vec{b}\left(\vec{k}\right)$
is the intrinsic spin-orbit field, with $\vec{b}\left(\vec{k}\right)=-\vec{b}\left(-\vec{k}\right)$
due to time reversal symmetry. Thus, for a three-dimensional system,
$\vec{b}$ can only be present if the inversion symmetry of the host
crystal is broken. In the case of a two-dimensional system, it is
conventional to talk about its two-dimensional bandstructure, and
to include the confinement potential in $\energyOfK{k}$ and $\vec{b}\left(\vec{k}\right)$
(instead of including it explicitly in $\Vslow$); in this case $\vec{b}$
can also result from an asymmetry in the confinement. 

In contrast, $\Hextrinsic$ does not require broken inversion symmetry
of the pure crystal or of the structure.  It is important to note
that $\SOcouplingPseudoSpin$ in Equation~(\ref{eq:Hextrinsic})
can be many orders of magnitude larger than the vacuum value $\SOcouplingVac$;
this is due to the large spin-orbit interaction when the Bloch electrons
move close to the nuclei, with velocities that are close to relativistic.
Both $\Hintrinsic$ and $\Hextrinsic$ may be important for the spin
Hall effect, as we will discuss in this article. We present specific
forms of these effective Hamiltonians in Secs.~\ref{sub:IntrinsicH}
and~\ref{sub:ExtrinsicH}.

\subsection{Band structure of materials with spin-orbit interaction}

\label{sub:Bandstructure}

We now consider the electron wavefunctions near the forbidden gap
of a semiconductor. These wave functions are often described by the
Kohn-Luttinger $\vec{k}\cdot\vec{p}$ method, where one expands the
Hamiltonian in terms of band edge Bloch functions. Here, we present
a brief overview of this method; more detailed explanations can be
found, e.g., in \possessivecite{BlountKP} review article or in the
books by \citeasnoun{BirPikusSymm} and by \citeasnoun{Winkler}.

\newcommand{\BlochKetKnotSymbol}[1]{u_{#1,0}}

\newcommand{\kdotpBasisSymbol}{\phi}
 For a given $\vec{k}$, the solutions of the Schr\"odinger equation
are Bloch functions $e^{i\vec{k}\cdot\vec{r}}u_{\nu',\vec{k}}\left(\vec{r}\right)$.
Here, $\nu'$ is the band index and includes the spin degree of freedom.
The lattice-periodic part $u_{\nu',\vec{k}}\left(\vec{r}\right)$
of these Bloch functions can be expanded in the functions $u_{\nu,\,\vec{k}=0}\left(\vec{r}\right)=\braket{\vec{r}}{\BlochKetKnotSymbol{\nu}}$,
which provide a complete basis when all bands $\nu$ are taken. 
For semiconductors with a direct gap at the center of the Brillouin
zone, which we discuss here, one may consider states in close vicinity
of $\vec{k}=0$, truncate this expansion, and only take the closest
bands into account. Therefore, it is sufficient to know the matrix
elements of the full Hamiltonian $H$ in the truncated basis $\ket{\kdotpBasisSymbol_{\nu}}=e^{i\vec{k}\cdot\vec{r}}\ket{\BlochKetKnotSymbol{\nu}}$,
i.e., one  considers $H_{\nu\nu'}\!\left(\vec{k}\right)=\braopket{\kdotpBasisSymbol_{\nu}}{H}{\kdotpBasisSymbol_{\nu'}}$.
More concretely, one evaluates \begin{equation}
H_{\nu\nu'}\!\left(\vec{k}\right)=E_{\nu}\delta_{\nu\nu'}+\frac{\hbar^{2}k^{2}}{2m_{0}}\delta_{\nu\nu'}+\frac{\hbar}{m_{0}}\vec{k}\cdot\bra{\BlochKetKnotSymbol{\nu}}\bpi\ket{\BlochKetKnotSymbol{\nu'}},\label{eq:Hkdotp}\end{equation}
where $E_{\nu}$ is the energy offset of the band at $\vec{k}=0$,
i.e., $\big[p^{2}/2m_{0}+\Vperiodic+\left(\SOcouplingVac/\hbar\right)\:\bsigma\cdot\left(\vec{p}\times\nabla\Vperiodic\right)\big]\ket{\BlochKetKnotSymbol{\nu}}=E_{\nu}\ket{\BlochKetKnotSymbol{\nu}}$.
Further, $\bpi=\vec{p}+\left(\SOcouplingVac/\hbar\right)\:\left(\nabla\Vperiodic\times\bsigma\right)$,
and one usually approximates $\bpi\approx\vec{p}$; then the last
term of Equation~(\ref{eq:Hkdotp}) is proportional to the matrix
element of $\vec{k}\cdot\vec{p}$, giving this method its name. Finally,
this finite-dimensional Hamiltonian $H\left(\vec{k}\right)$ now describes
the band structure in terms of a few parameters---band offsets and
momentum matrix elements of $\vec{k}=0$ Bloch functions---and is
well suited for analyzing charge carriers. 

\begin{figure}
\begin{center}\includegraphics[height=50mm,keepaspectratio]{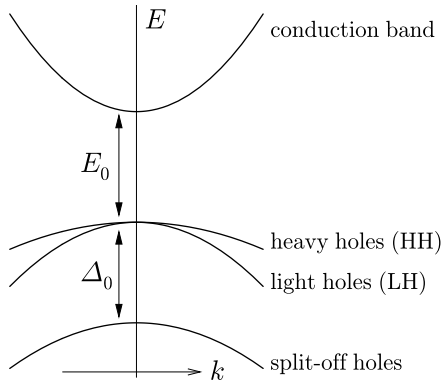}\end{center}

\caption{\label{cap:BandStructure}Schematic band structure of a cubic direct
gap semiconductor. When spin-orbit interaction is disregarded, one
finds an \emph{s}-like conduction band and a \emph{p}-like three-fold
degenerate valence band. The spin-orbit interaction due to crystal
potential $\Vperiodic$ {[}entering as $\Vtotal$ in Equation (\ref{eq:HVacuum}){]}
partially lifts this degeneracy and leads to a substantial splitting
between the valence bands with total angular momentum $\JabsHoles$
(heavy and light holes) and those with $\Jabs=1/2$ (split-off holes).}
\end{figure}

Alternatively, one can use a second method and construct a Hamiltonian
by allowing all contributions (up to some order in $k$) that are
invariant under the symmetry operations of the system---the coupling
constants are material-dependent parameters. For example, when considering
the top of the valence band in Fig.~\ref{cap:BandStructure} and
in the absence of inversion asymmetry, magnetic field, and strain,
the most general form up to quadratic terms in $k$ (i.e., in the
effective mass approximation), is the $4\times4$ Luttinger Hamiltonian,
\begin{equation}
H_{\mathrm{L}}=\frac{\hbar^{2}}{m_{0}}\left[\left(\gamma_{1}+\frac{5}{2}\gamma_{2}\right)\frac{k^{2}}{2}-\gamma_{3}\left(\vec{k}\cdot\vec{J}\right)^{2}+\left(\gamma_{3}-\gamma_{2}\right)\sum\nolimits _{i}k_{i}^{2}J_{i}^{2}\,\right],\label{eq:Luttinger}\end{equation}
which is consistent with the cubic symmetry. Here, $\vec{J}=\left(J_{x},\, J_{y},\, J_{z}\right)$
and $J_{i}$ are the angular momentum matrices for spin $\frac{3}{2}$,
and $\gamma_{i}$ are the material-dependent Luttinger parameters.
 $H_{\mathrm{L}}$ describes $p$-doped Si and Ge; for GaAs, due
to broken inversion symmetry, terms linear in $k$ arise as well. 

 By contrast, using the first method ($\vec{k}\cdot\vec{p}$ method)
instead, the Hamiltonian $H_{\nu\nu'}\!\left(\vec{k}\right)$ {[}Equation~(\ref{eq:Hkdotp}){]}
is evaluated directly for the eight bands $\nu$ (including spin)
shown in Fig.~\ref{cap:BandStructure}, and one arrives at the simplest
version of the $8\times8$ Kane model. It only includes three parameters,
namely, the energy gap $\Egap$, the energy of the split-off holes
$\DeltaSO$, and the matrix element $P$ of the momentum (multiplied
by $\hbar/m_{0}$) between \emph{s}- and \emph{p}-type states. $P$
is nearly universal for III-V compounds, while the other parameters
depend on the material. This eight-dimensional description is accurate
for narrow band materials; for wider gaps it still provides understanding
at a qualitative level.

Furthermore, one can then derive an effective, lower-dimensional Hamiltonian
by block-diagonalizing $H_{\nu\nu'}$, this is an efficient way to
calculate the band structure in the vicinity of $\vec{k}=0$. One
can do this either exactly or by using time-independent degenerate
perturbation theory (see Sec.~\ref{sub:Effective-Hamiltonian}).
Considering a particular block, this allows estimating the magnitude
of the symmetry-allowed terms. Terms that were not present before
block-diagonalization are called contributions from remote bands.
For example, using the $8\times8$ Kane model, one can calculate the
parameters $\gamma_{i}$ that enter the $4\times4$ Luttinger Hamiltonian
{[}Equation~(\ref{eq:Luttinger}){]} for the top of the valence band;
because the model is isotropic, one gets $\gamma_{2}=\gamma_{3}$.
To estimate corrections due to the cubic symmetry, one needs to take
more bands into account.

In addition to the crystal field, we also wish to consider electric
fields that are applied externally or that result from charged impurities.
Assuming that the corresponding potential $\Vslow\left(\vec{r}\right)$
varies slowly on the scale of a lattice constant, we can apply the
envelope function approximation (EFA), i.e., we replace the plain
waves $e^{i\vec{k}\cdot\vec{r}}$ by slowly varying envelope functions
$\psi_{\nu}\left(\vec{r}\right)$. Evaluating matrix elements of $\Vslow$
in the basis $\braket{\vec{r}}{\psi,\nu}=\psi_{\nu}\left(\vec{r}\right)u_{\nu,\,\vec{k}=0}\left(\vec{r}\right)$,
the main contribution is diagonal with respect to the band index $\nu$.
However,  because $\vec{k}$ and $\Vslow\left(\vec{r}\right)$ do
not commute, off-diagonal elements (in $\nu$) can arise in the Hamiltonian
when expanding in $\vec{k}$, see e.g. Equation~(\ref{eq:Hextrinsic}).
 Magnetic fields can be included in a similar way, using the Perierls
substitution $\hbar\vec{k}=-i\hbar\nabla-\left(e/c\right)\vec{A}$.
Here, we use $e<0$ for electrons and $e>0$ for holes. Then, the
components of $\vec{k}$ no longer commute, $\vec{k}\times\vec{k}=i\left(e/\hbar c\right)\vec{B}$;
this leads to Zeeman coupling, which is described by an effective
$g$ factor.

\subsection{Effective Hamiltonian}

\label{sub:Effective-Hamiltonian}

The $\vec{k}\cdot\vec{p}$ method leads to high-dimensional Hamiltonians
{[}Equation~(\ref{eq:Hkdotp}){]}, for example, an $8\times8$ matrix
for the Kane model, thus further simplifications are desirable. 
For this, one can use time-independent degenerate perturbation theory
and describe a subset of states (say, the lowest conduction band states)
with an effective Hamiltonian. L\"owdin partitioning is a straightforward
and convenient method to implement such a perturbative expansion \cite{Winkler};
it is also known as Foldy-Wouthuysen transformation in the context
of the Dirac equation and as Schrieffer-Wolf transformation in the
context of the Anderson model. The idea is to find a unitary transformation
$e^{-S}$ (i.e., $S$ is anti-Hermitian) such that the transformed
Hamiltonian $e^{-S}He^{S}$ is block diagonal, i.e., contains no off-diagonal
elements between the states we are interested in and any other states.
This procedure assumes that the states of interest (e.g., the conduction
band) are separated from the other states (all other bands) by an
energy much larger than the Fermi energy. Then, because these off-diagonal
elements are small, one can eliminate the off-diagonal blocks of $e^{-S}He^{S}$
order by order (or even exactly). In our example, the transformed
Hamiltonian consists of one $2\times2$ and one $6\times6$ block.
The smaller block describes the conduction band electrons---we can
understand the two dimensions as (pseudo-) spin $\frac{1}{2}$. At
the level of wave functions, the periodic part of the electron wave
function at a given $\vec{k}\neq0$ is mainly described by the conduction
band Bloch function at $k=0$, but also contains a small admixture
from the valence band Bloch functions.

\subsection{Intrinsic spin-orbit coupling}

\label{sub:IntrinsicH}One generally distinguishes between intrinsic
and extrinsic mechanisms of spin-orbit coupling; however, this classification
is not unique across the literature. In this article, we classify
it according to the individual terms of the effective Hamiltonians.
Namely, we refer to the spin-orbit contributions to the Hamiltonian
that depend on impurity potentials as \emph{extrinsic}. The other
spin-orbit contributions arise even in the absence of impurities and
we call them \emph{intrinsic}---we also call effects resulting from
these contributions intrinsic, even if we must allow for a small concentration
of impurities to make the theory of dc transport properties consistent.%
\footnote{This is reminiscent of the definition of an {}``intrinsic semiconductor,''
which is so pure that (at a sufficiently high temperature) the impurity
contribution to the carrier density is negligible. The conductivity
of such a sample is known as \emph{intrinsic} conductivity (again,
a finite transport lifetime $\tau$ is required to make the conductivity
well-defined). At lower temperatures, the carrier density mainly results
from the impurities and now one refers to \emph{extrinsic} properties. %
} 

For a (pseudo-) spin-$\frac{1}{2}$ system, the spin-orbit part of
the intrinsic one-particle Hamiltonian has the general form $\Hintrinsic$
{[}Equation~(\ref{eq:HIntrinsic}){]}. In the following, we discuss
the origin and the functional form of such spin-orbit fields. We focus
on such spin-$\frac{1}{2}$ descriptions, because they are relevant
for low-dimensional systems and are the basis of most theoretical
works. 

We first consider a $n$-doped bulk (3D) semiconductor and the effective
Hamiltonian of conduction band electrons. III-V and II-VI semiconductors
lack inversion symmetry and are available in two modifications: in
cubic zinc blende or in hexagonal wurtzite structure. In zinc blende
modification, the bulk inversion asymmetry (BIA) leads to the Dresselhaus
term \begin{equation}
\HDresselhausThreeD=\couplingDresselhausThreeD\, k_{x}\left(k_{y}^{2}-k_{z}^{2}\right)\sigma_{x}+\mathrm{c.p.},\label{eq:HDresselhausThreeD}\end{equation}
where $k_{i}$ are along the principal crystal axes. Here, $\mathrm{c.p.}$
stands for cyclic permutation of all indices, and the symmetrized
product of the components $k_{i}$ must be used if a magnetic field
is applied. The Dresselhaus term originates from bands further away
than the basic eight bands, and one finds the coupling constant in
terms of the band parameters; when using the extended $14\times14$
Kane model its numerical value is $\couplingDresselhausThreeD\approx27\:\mathrm{eV}\Angstrom{}^{3}$,
for both GaAs and InAs \cite{Winkler}. However, tight-binding calculations
and interpretation of weak-localization experiments indicate lower
values, at least for GaAs \cite{PhysRevB_53_3912,Krichk3}.

When the electrons are confined to two dimensions, the expectation
value of the Dresselhaus term along the confinement direction (that
we always assume to be along {[}001{]}) should be taken, $\left\langle \HDresselhausThreeD\right\rangle $.
While $\left\langle k_{z}\right\rangle =0$, we see that the terms
in $\left\langle k_{z}^{2}\right\rangle \approx\left(\pi/d\right)^{2}$
are large for small confinement width $d$, thus the main BIA contribution
becomes \begin{equation}
\HDresselhausk=\beta\left(k_{x}\sigma_{x}-k_{y}\sigma_{y}\right),\label{eq:HDresselhausk}\end{equation}
with $\beta\approx-\couplingDresselhausThreeD\,\left(\pi/d\right)^{2}$.
In addition to the $k$-linear term in Equation (\ref{eq:HDresselhausk}),
there is also a $k^{3}$-term,\begin{equation}
\HDresselhausTwoDkkk=\couplingDresselhausThreeD\, k_{x}k_{y}\left(k_{y}\sigma_{x}-k_{x}\sigma_{y}\right),\label{eq:HDresselhaus2dk3}\end{equation}
which is small compared to $\HDresselhausk$ in the strong confinement
(low carrier density) limit $\pi/d\gg k_{\mathrm{F}}$, where $k_{\mathrm{F}}$
is the Fermi wave vector. Additionally, a spin-orbit coupling term
arises if the confinement potential $V\left(z\right)$ along the $z$-direction
is not symmetric, i.e., if there is a structure inversion asymmetry
(SIA). Equation~(\ref{eq:Hextrinsic}) provides an explicit connection
between such a potential and spin-orbit coupling. Taking the expectation
value $\left\langle \Hextrinsic\right\rangle $ along the $z$-direction
and noting that the only contribution of the confinement field is
$\propto\left\langle \nabla_{z}\Vslow\right\rangle $, one finds the
Rashba Hamiltonian,\begin{equation}
\HRashbaEl=\alpha\left(k_{y}\sigma_{x}-k_{x}\sigma_{y}\right),\label{eq:RashbaEl}\end{equation}
corresponding to $\vec{b}\left(\vec{k}\right)=2\alpha\:\hat{\vec{z}}\times\vec{k}$.
 More generally, for spinors with $J_{z}=\pm1/2$, $\HRashbaEl$
is the only $k$-linear invariant of the group $C_{\infty v}$ that
takes into account the confinement potential $\Vslow(z)$ but disregards
the discrete symmetry of the crystal. The magnitude of the coupling
constant $\alpha$ depends on the confining potential and it can be
modified by applying an additional field via external gates. It also
defines the spin-precession wave vector $k_{\alpha}=\alpha m/\hbar^{2}$.
Finally, such a term $\HRashbaEl$ is also present for three-dimensional
electrons in systems of hexagonal wurtzite structure (or in cubic
systems with strain, see Sec\@.~\ref{sub:Strain}).

Next, we consider a $p$-doped three dimensional semiconductor, i.e.,
the $\JabsHoles$ valence band, described by the four-dimensional
Luttinger Hamiltonian. Remote bands lead to a BIA contribution to
the Hamiltonian, which is given by Equation (\ref{eq:HDresselhausThreeD})
after replacing $\sigma_{i}$ by the angular momentum matrices $J_{i}$
for spin $\frac{3}{2}$ and using a different coupling constant. If
the system is reduced to two dimensions, size quantization lifts the
fourfold degeneracy at $\vec{k=0}$ and creates heavy-hole (HH) bands
with $\Jz=\pm3/2$, and light-hole (LH) bands with $\Jz=\pm1/2$ (for
confinement along the $[001]$ axis). Usually the HH bands are higher
in energy, thus for small doping it is sufficient to consider only
them.  For spinors with $J_{z}=\pm3/2$, the only invariant of
the group $C_{\infty v}$ (again, we do not discuss invariants of
discrete symmetries here) respecting time reversal symmetry is\begin{equation}
\HRashbaHTwoD=i\alpha_{\mathrm{h}}\left(k_{-}^{3}\sigma_{+}-k_{+}^{3}\sigma_{-}\right),\label{eq:HRashbaHTwoD}\end{equation}
where $a_{\pm}\equiv a_{x}\pm ia_{y}$ for any $a$. As distinct from
$\HRashbaEl$ {[}Equation~(\ref{eq:RashbaEl}){]}, the Rashba Hamiltonian
for heavy holes is cubic in $k$, as it was discussed in \cite{Winkler2DHoles,Schliemann_2DHolesk3}.

\subsection{Extrinsic spin-orbit coupling}

\label{sub:ExtrinsicH}
\newcommand{\Vhole}{V_{\mathrm{h}}}
Electric fields due to impurities lead to extrinsic contributions
to the spin-orbit coupling. Externally applied electrical fields lead
to analogous contributions. To derive the dominant extrinsic term,
it is sufficient to restrict ourselves to the simplest $8\times8$
Kane Hamiltonian; higher bands will give rise to small corrections.
Using third-order perturbation theory and for conduction band electrons,
we find $\Hextrinsic$ as given in Equation~(\ref{eq:Hextrinsic}),
with \cite{Nozieres73,Winkler}

\begin{equation}
\SOcoupling\approx\frac{P^{2}}{3}\left[\frac{1}{\Egap^{2}}-\frac{1}{\left(\Egap+\DeltaSO\right)^{2}}\right],\label{eq:HextrinsicEl}\end{equation}
and where $V$ is the potential due to impurities and an externally
applied field. It is noteworthy that Equation (\ref{eq:Hextrinsic})
has the same analytical form as the vacuum spin-orbit coupling {[}Equation
(\ref{eq:HVacuum}){]}; this is because both the Dirac equation and
the simplest Kane Hamiltonian have spherical symmetry and because
both the Pauli equation and Equation~(\ref{eq:Hextrinsic}) are obtained
in a low-energy expansion. However, for $\DeltaSO>0$ the couping
constant $\SOcoupling$ has the \emph{opposite sign} as in vacuum.

One finds $\SOcouplingEl\approx5.3\:\mbox{\AA}{}^{2}$ for GaAs and
$\SOcouplingEl\approx120\:\mbox{\AA}{}^{2}$ for InAs, i.e., spin-orbit
coupling in $n$-GaAs is by \textit{\emph{six orders of magnitude
stronger}} than in vacuum, and even larger for InAs due to its smaller
gap. This enhancement of spin-orbit coupling is critical for developing
large extrinsic spin currents. Furthermore, for a two-dimensional
system, when considering $V$ as averaged along the $\hat{\vec{z}}$-direction,
both $\nabla\Vslow$ and $\vec{k}$ are in-plane, thus we have $\HextrinsicEl=\SOcouplingEl\:\sigma_{z}\left(\vec{k}\times\nabla\Vslow\right)_{z}$.

For a 3D hole system, we consider the $\JabsHoles$ valence band.
Then, the dominant extrinsic spin-orbit term in third order perturbation
theory describing the valence band states is\begin{equation}
\HextrinsicVal=\SOcouplingVal\,\vec{J}\cdot\left(\vec{k}\times\nabla\Vslow\right),\label{eq:HextrinsicHole}\end{equation}
with $\SOcouplingVal=-P^{2}/3\Egap^{2}$, i.e., for GaAs $\SOcouplingVal\approx-15\,\mbox{\AA}^{2}$
\cite{Winkler}. and has to be added to the Luttinger Hamiltonian
$H_{\mathrm{L}}$ {[}Equation~(\ref{eq:Luttinger}){]}.  When considering
a two-dimensional hole system with HH-LH splitting, we can restrict
Equation (\ref{eq:HextrinsicHole}) to the heavy holes states, where
$\Jz=\pm3/2$. Expressing this two-dimensional subspace in terms of
a pseudo-spin $\frac{1}{2}$ leads to\begin{equation}
\HextrinsicVal=-\frac{P^{2}}{2\Egap^{2}}\;\left(\vec{k}\times\nabla\Vslow\right)_{z}\sigma_{z}.\label{eq:HextrinsicHoles2D}\end{equation}
Thus, the extrinsic spin-orbit interaction for two-dimensional heavy
hole states has the same form as for two-dimensional electrons.

Finally we point out that extrinsic spin-orbit coupling arises because
the the long range Coulomb potential of the impurities does not commute
with the intrinsic Hamiltonian of the hosting crystal. The extrinsic
Hamiltonian $\Hextrinsic$ {[}Equations~(\ref{eq:Hextrinsic}) and
(\ref{eq:HextrinsicHole}){]} is obtained in the framework of the
EFA, which disregards short-range contributions to the spin-orbit
coupling arising from the chemical properties of dopants. This is
why the coupling $\SOcoupling$ depends only on the parameters of
the perfect crystal lattice.

\subsection{Strain}

\label{sub:Strain} Non-hydrostatic strain reduces the symmetry of
the system and in this way leads to additional spin-orbit terms in
the Hamiltonian. In third order perturbation theory of the Kane Hamiltonian,
the effective conduction-band Hamiltonian due to strain is dominated
by \begin{equation}
\HstrainEl=\frac{-2C_{2}\DeltaSO P}{3\Egap(\Egap+\DeltaSO)}\,\bsigma\cdot\left(\vec{k}\times\bvarepsilon_{s}\right)\equiv\frac{1}{2}C_{3}\,\bsigma\cdot\left(\vec{k}\times\bvarepsilon_{s}\right),\label{eq:HstrainEl}\end{equation}
where $\bvarepsilon_{s}=\left(\varepsilon_{yz},\:\varepsilon_{xz},\:\varepsilon_{xy}\right)$
describes the shear strain. Here, $C_{2}$ is the interband deformation-potential
constant that arises in noncentrosymmetric semiconductors \cite{Winkler,Trebin_Resonances}.
Note that \citeasnoun{PikusTitkov} as well as \citeasnoun{IvchenkoPikus}
use the opposite sign in the definition of this constant, $C_{2}^{\mathrm{PT/IP}}=-C_{2}$.
Further, if a shear is applied such that only $\varepsilon_{xy}\neq0$,
Equation~(\ref{eq:HstrainEl}) has the same form as the Rashba Hamiltonian
{[}Eq.~(\ref{eq:RashbaEl}){]}.

For three-dimensional $\JabsHoles$ valence band states, the main
strain contribution is \cite{PikusTitkov} \begin{equation}
\HstrainVal=\frac{2C_{2}P}{3\Egap}\,\vec{J}\cdot\left(\vec{k}\times\bvarepsilon_{s}\right).\label{eq:HstrainHole}\end{equation}
Note that when the system is confined to two dimensions, Equation~(\ref{eq:HstrainHole})
inplies a $k$-linear spin-orbit contribution for heavy hole states
due to strain, $(C_{2}P/\Egap)\,\left(\vec{k}\times\bvarepsilon_{s}\right)_{z}\sigma_{z}$.
This linear term arises due to the low symmetry of the strained material,
in contrast to the Rashba term {[}cf.\  Equation (\ref{eq:HRashbaHTwoD}){]},
which is dominated by terms cubic in $k$ and where the $k$-linear
terms are numerically small \cite{Winkler}.

\subsection{Anomalous velocity and coordinate }

\newcommand{\anomalousCoordinate}{\delta\vec{r}}
\label{sub:Anomalous-velocity}As a consequence of the spin-orbit
interaction, velocity and coordinate operators are modified and become
spin dependent---this will be important when considering currents.
When an effective Hamiltonian is derived in perturbation theory, as
explained in Sec.~\ref{sub:Effective-Hamiltonian}, a unitary transformation
$e^{-S}$ is applied. Thus, the coordinate operator $\vec{r}=i\left(\partial/\partial\vec{k}\right)$
is also transformed, $\vec{r}\mapsto\tilde{\vec{r}}=e^{-S}\vec{r}e^{S}=\vec{r}+\anomalousCoordinate$
and we call $\anomalousCoordinate$ the anomalous coordinate. In particular,
because $S$ couples to spin, $\anomalousCoordinate$ is spin-dependent.
This correction $\delta\vec{r}$ is known as the Yafet term \cite{YafetSSP};
also, it can be expressed in terms of a Berry connection, $\braopket{u_{\nu',\,\vec{k}}}{i\nabla_{\vec{k}}}{u_{\nu,\,\vec{k}}}$
\cite{AnnPhysNy_160_343,SundaramNiu_WavePackets,NagaosaAHE06}. In
perturbation theory, one finds \begin{align}
\delta\vec{r}_{\mathrm{SO,\, e}} & =\SOcouplingEl\left(\bsigma\times\vec{k}\right),\label{eq:CoordEl}\\
\delta\vec{r}_{\mathrm{SO,\, v}} & =\SOcouplingVal\left(\vec{J}\times\vec{k}\right),\label{eq:CoodHole}\end{align}
for conduction band electrons and for $\JabsHoles$ heavy hole states,
respectively. Note that coordinate operators no longer commute, $\vec{r}\times\vec{r}=i\SOcouplingEl\bsigma$
and $\vec{r}\times\vec{r}=i\SOcouplingVal\vec{J}$, resp. Finally,
$\delta\vec{r}_{\mathrm{SO}}$ leads to an extra term in the equations
of motion that can be understood as anomalous velocity \cite{BlountKP}.

Formally, we can derive the anomalous velocity similarly to the coordinate,
namely $\vec{v}\mapsto e^{-S}\vec{v}e^{S}=\vec{v}_{0}+\delta\vec{v}$,
where $\delta\vec{v}$ is the anomalous velocity and, for a parabolic
band, $\vec{v}_{0}=\hbar\vec{k}/m^{*}$. Alternatively, one can obtain
the velocity operator from the Heisenberg equation, $\vec{v}=\left(i/\hbar\right)\left[H,\,\tilde{\vec{r}}\right]$.
For $H=H_{0}+H_{\mathrm{SO}}$, where $H_{\mathrm{SO}}$ contains
the (small) spin-orbit coupling, we get $\vec{v}=\vec{v}_{0}+\left(i/\hbar\right)\left[H_{\mathrm{SO}},\,\vec{r}\right]+\left(i/\hbar\right)\left[H_{0},\,\delta\vec{r}_{\mathrm{SO}}\right]$.
Thus, $H_{\mathrm{SO}}$ leads to an anomalous velocity because it
does not commute with the unperturbed coordinate $\vec{r}$, and,
additionally, the contribution from the anomalous coordinate $\delta\vec{r}_{\mathrm{SO}}$
should also be taken into account, as it can be significant. 

When the impurity potential $\Vslow$ is included, the above argument
remains the same, but now $H_{\mathrm{SO}}$ contains the extrinsic
contribution as well. Note that for extrinsic spin-orbit $H_{\mathrm{SO}}=\HextrinsicEl$
{[}Equation~(\ref{eq:Hextrinsic}){]}  the commutator $\left[\HextrinsicEl,\,\vec{r}\right]$
and the anomalous coordinate $\delta\vec{r}_{\mathrm{SO}}$ give equal
contributions to $\delta\vec{v}$ \cite{Nozieres73}.

\section{Mechanisms of spin transport}

\label{sec:SO-Mechanisms}We now address how the microscopic mechnism
of spin-orbit interaction, given by effective Hamiltonians, influences
spin transport and accumulation. In the following, we assume a non-interacting
system in the absence of a magnetic field. Because we ignore electron-electron
interaction, we do not consider the spin-drag effect here \cite{HVspinDrag},
which can lead to a suppression of spin transport at high temperatures,
and suppression of spin-relaxation \cite{GlazovIvchenkoSpRelax}.
We also restrict ourselves to Boltzmann transport and do not discuss
the hopping regime \cite{SHE_hopping}.

\subsection{Intrinsic: spin precession}

\label{sub:Intrinsic-Spin-Precession}In a system with weak intrinsic
spin-orbit coupling $\Hintrinsic$ {[}Equation~(\ref{eq:HIntrinsic}){]},
consider a carrier with spin aligned along the spin-orbit field $\vec{b}\left(\vec{k}\right)$.
When an electrical field $\vec{E}=E\hat{\vec{x}}$ is applied, the
particle is accelerated: $\dotvec{k}=e\vec{E}/\hbar$ in lowest order
in spin orbit-interaction; and its spin-orbit field changes: $\dotvec{b}=\left(\partial\vec{b}/\partial k_{x}\right)eE/\hbar$.
For a small acceleration, the spin follows adiabatically the direction
of $\vec{b}\left(\vec{k}\right)$. Additionaly, there will be a non-adiabatic
correction that can be derived as follows. Say, the direction of $\vec{b}$
rotates in the $xy$ plane (as it is the case for Rashba interaction
$\HRashbaEl$). Because the rotation frequency $\omega$ is the component
of $\dotvec{b}/b$ perpendicular to $\vec{b}$, it is $\omega=(\vec{b}\times\dotvec{b})_{z}/b^{2}$.
In the co-rotating frame, there is a field $b$ along the $x$-axis
and a field $\hbar\omega$ along the $z$-axis. As we are interested
in the next-to-adiabatic correction, we assume that $\omega$ changes
slowly and that the spin remains aligned along the total field in
the rotating frame, thus it has a component $s_{z}\approx\left(\hbar/2\right)\hbar\omega/b$.
Therefore the first non-adiabatic correction to the spin is $\delta\vec{s}\left(\vec{k}\right)=\hbar^{2}(\vec{b}\times\dotvec{b})_{z}/2b^{3}$.
In particular, the electrical field drives this spin precession (via
$\dotvec{b}$), leading to such out-of-plane component $\delta\vec{s}\left(\vec{k}\right)$,
which could be important in spin transport \cite{Sinova04}, see Sec.~\ref{sub:Bulk-Spin-Currents}.
However, when considering dc properties, one must be careful and also
allow for impurities that decelerate the carriers to reach a steady
state. In particular, if $\vec{b}\left(\vec{k}\right)$ is linear
in $\vec{k}$, it turns out that the deceleration at impurities cancels
this spin precession, see Sec.~\ref{sub:Bulk-Spin-Currents}.

\subsection{Extrinsic: skew scattering}

\label{sub:Skew-Scattering}
\begin{figure}
\begin{center}\includegraphics[%
  height=30mm,
  keepaspectratio]{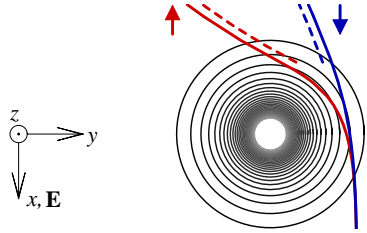}\end{center}

\caption{\label{cap:skew} Spin-dependent scattering of electrons at an attractive
impurity. We show the classical trajectories (solid lines), for a
screened Coulomb potential and for strongly exaggerated extrinsic
spin-orbit coupling {[}using Equation (\ref{eq:Hextrinsic}) with
$\SOcouplingEl>0${]} and with spin quantization axis perpendicular
to the plane. The skew-scattering current results from different scattering
angles for spin-$\uparrow$ and spin-$\downarrow$ electrons and leads
to a positive spin Hall conductivity, $\sigmaSHSS=-\jSHc_{\mathrm{SS},y}^{z}/E_{x}>0$.
 The dashed lines show the horizontal displacement due to the side
jump effect, contributing to the spin current with opposite sign.
}
\end{figure}

When a carrier scatters at an impurity
potential $V$, because of the extrinsic spin-orbit interaction {[}Equations
(\ref{eq:Hextrinsic}) and (\ref{eq:HextrinsicHoles2D}){]} the scattering
cross section depends on the spin state \cite{SmitSkew}, see Fig.~\ref{cap:skew}.
This effect is known as Mott skew scattering \cite{MottMassey} and
was originally considered for high-energy electrons that are elastically
scattered by an atom and that are described by the vacuum Hamiltonian,
Equation (\ref{eq:HVacuum}). Skew scattering does not appear in the
first order Born approximation, thus is at least of the order $\Vslow^{3}$.
For band electrons, skew scattering was originally considered as the
origin of the anomalous Hall effect, see Sec.~\ref{sub:AHE}. As
applied to spin Hall effct, the relevance of this extrinsic mechanism
was recognized early on \cite{DPpolarization,Hirsch99,Zhang00}.

\subsection{Extrinsic: side jump mechanism}

\label{sub:Side-Jump-Mechanism}The side jump mechanism \cite{BergerSideJump}
describes the lateral displacement of the wave function during the
scattering event. (Such a displacement does not modify the skew scattering
cross section introduced in Sec.~\ref{sub:Skew-Scattering}, because
it does not change the scattering angle measured at large distances.)
The side-jump contribution is obtained when the anomalous velocity
$\delta\vec{v}$ (see Sec. \ref{sub:Anomalous-velocity}) is integrated
over the duration of the scattering process. As indicated in Sec.~\ref{sub:Anomalous-velocity},
the anomalous coordinate for electrons {[}Equation (\ref{eq:CoordEl}){]}
leads to an anomalous velocity contribution $\SOcouplingEl(\bsigma\times\dotvec{k})$;
and there is an equal term due to $(i/\hbar)[\HextrinsicEl,\:\vec{r}]$.
For impurity scattering with momentum transfer $\delta\vec{k}$, this
results in a total lateral displacement $2\lambda(\bsigma\times\delta\vec{k})$
{[}and anologousely for holes, using Equation (\ref{eq:CoodHole}){]}.
When also the effect of the anomalous velocity due to an applied electrical
field is considered, the side-jump contribution to spin-transport
becomes a subtle issue, for a detailed analysis and intuitive description
see \cite{Nozieres73}. Because the side jump mechanism is not contained
in the Boltzmann approach, in such a framework it needs to be evaluated
separately. The side jump contribution can be found using the Kubo
formula and diagrammatic approaches; see \cite{LewinerDiagramms,CrepieuxBrunoAHE,TseDasSarmaExtrinsic}.

\subsection{Kinetic equation}

\label{sub:Kinetic-equations}
\newcommand{\fDiagonal}{n}

\newcommand{\fSpin}{\vec{f}}

\newcommand{\velocityE}{v_{\epsilon}}

\newcommand{\azimuthalAngle}{\varphi}

\newcommand{\Vimpurity}{V_{\mathrm{i}}}

\newcommand{\fMatrixE}{\hat{\Phi}}

\newcommand{\fSpinE}{\boldsymbol{\Phi}}

\newcommand{\fDiagonalE}{\Phi_{\mathrm{c}}}

We consider the Hamiltonian $\Heff$ {[}Equation~(\ref{eq:Heffective}){]}
containing both intrinsic ($\Hintrinsic$) and extrinsic ($\Hextrinsic$)
spin-orbit contributions, and with $\Vslow$ that describes the electrical
field $\vec{E}$ and the impurity potential $\Vimpurity$. For a homogeneous
system of non-interacting particles, one can derive the kinetic equation
\cite{Khaetskii_2DGeneralized,ShytovKinetic} 

\begin{equation}
\frac{\partial\hat{f}}{\partial t}+\frac{1}{\hbar}\,\bsigma\cdot\left(\vec{b}\times\fSpin\right)+e\vec{E}\cdot\frac{1}{\hbar}\frac{\partial\hat{f}_{0}}{\partial\vec{k}}=\left(\frac{\partial\hat{f}}{\partial t}\right)_{\mathrm{coll.}},\label{eq:Kinetic}\end{equation}
i.e., a spin-dependent Boltzmann equation, where the distribution
function is written as a $2\times2$ spin matrix $\hat{f}=\hat{f}_{0}(\vec{k})+\frac{1}{2}\fDiagonal(\vec{k})\,1\!\!1+\fSpin(\vec{k})\cdot\bsigma$,
with equilibrium distribution function $\hat{f}_{0}$. Here, $n$
is the excess particle density and $\fSpin$ describes the spin polarization
density.  Formally, the Boltzmann equation is obtained by an expansion
in $1/k_{\mathrm{F}}\ell$, where $\ell$ is the mean free path, i.e.,
it is applicable for dilute impurities. Traditionally, the Boltzman
equation describes the distribution function $\fDiagonal(\vec{k})$,
which is the \emph{probability} density of a state $\vec{k}$ to be
occupied---in contrast, here it describes $\hat{f}(\vec{k})$, which
corresponds to the $2\times2$ \emph{density matrix} for a spin-$\frac{1}{2}$
particle.

All terms on the l.h.s. of Equation~(\ref{eq:Kinetic}) arise in
the absence of impurities. The first term is the derivative of $\hat{f}$
with respect to its explicit time-dependence.  The second term describes
the spin precession; it is obtained from the Heisenberg equation,
$\fSpin\cdot\dotvec{\bsigma}=\fSpin\cdot\left(i/\hbar\right)\big[-\frac{1}{2}\vec{b}\left(\vec{k}\right)\cdot\bsigma,\:\bsigma\big]$.
The third term is the driving term due to the electrical field, given
in lowest order of $\vec{E}$. Finally, for inhomogeneous particle
and spin distributions, the term $\vec{v}\cdot\nabla\hat{f}$ has
to be added to the l.h.s.

The r.h.s.\ of the Boltzmann equation {[}Equation~(\ref{eq:Kinetic}){]}
is the collision term, symbollically given by\begin{equation}
\left(\frac{\partial\hat{f}(\vec{k})}{\partial t}\right)_{\mathrm{coll}}=n_{i}v\sum_{\vec{k}';\:\,\epsilon'=\epsilon}\,\frac{\crossSecSpin}{d\Omega}\,\left[\hat{f}(\vec{k}')-\hat{f}(\vec{k})\right],\label{eqp:collTerm}\end{equation}
 Here, $n_{i}$ is the impurity density and we only consider elastic
scattering $\vec{k}\to\vec{k}'$ and $\vec{k}'\to\vec{k}$.  The
scattering cross section tensor $\crossSecSpin/d\Omega$ and the summation
over final states $\vec{k}'$ are spin-dependent because (i) the extrinsic
interaction $\Hextrinsic$ leads to skew-scattering, (ii) the intrinsic
spin-orbit Hamiltonian $\Hintrinsic$ induces a spin-dependent density
of states (DOS), and (iii) $\Hintrinsic$ causes spin-dependent momentum
transfer---in general, the spin dependence of scattering is rather
complex. The description of scattering is simplified for weak spin
orbit coupling, as the collision term can be expanded in spin-orbit
coupling, and  we can discuss the individual corrections seperately.
Note that the Boltzmann equation does not include the side-jump effect
(cf.\ Sec.~\ref{sub:Side-Jump-Mechanism}), which is of higher order
in $1/k_{\mathrm{F}}\ell$.

When considering only spin-orbit coupling due to $\Hextrinsic$, the
collision term including skew scattering for a central symmetric impurity
potential is \cite{EHR_extrinsic}\begin{equation}
\left(\frac{\partial\hat{f}(\vec{k})}{\partial t}\right)_{\mathrm{coll}}=n_{i}v_{\epsilon}\int d\Omega\!\left(\vec{k}'\right)\:\bigg\{ I(\vartheta)\left[\hat{f}(\vec{k}')-\hat{f}(\vec{k})\right]+\frac{1}{2}I(\vartheta)S(\vartheta)\,\bsigma\cdot\vec{m}\:\left[\fDiagonal(\vec{k})+\fDiagonal(\vec{k}')\right]\bigg\}\label{eq:CrossSectionExtrinsic}\end{equation}
 where $\vec{m}=\vec{k}'\times\vec{k}/\left|\vec{k}'\times\vec{k}\right|$
is the unit vector normal to the scattering plane and $\vartheta=\vartheta_{\vec{k}\vec{k'}}$
is the angle between $\vec{k}'$ and $\vec{k}$. The coefficient $I(\vartheta)$
is the spin-independent part of the scattering cross section, while
$S(\vartheta)$ is the so-called Sherman function \cite{MottMassey,MotzRMP,VoskoboynikovPRB},
describing the polarization of outgoing particles (which is normal
to the scattering plane) scattered into direction $\vec{k}$ from
an unpolarized incoming beam of momentum $\vec{k}'$. Note that $I\, S$,
as mentioned earlier in Sec.~\ref{sub:Skew-Scattering}, vanishes
in first-order Born approximation; the lowest term is $\propto V^{3}$.
Also, $S$ is proportional to spin-orbit coupling, this is why in
the second term in Equation~(\ref{eq:CrossSectionExtrinsic}), only
the spin-independent part of $\hat{f}$, $\frac{1}{2}\fDiagonal$,
is retained.

So far, we considered the distribution function $\hat{f}\left(\vec{k}\right)$
as density in $\vec{k}$-space. In the presence of intrinsic spin-orbit
interaction, the energy spectrum contains two branches: for a given
$\vec{k}$, there are two energies, split by the intrinsic field $\vec{b}\left(\vec{k}\right)$.
Thus, for elastic scattering, energy is conserved but $\left|\vec{k}\right|$
is not. It is now more convenient to choose a distribution as function
of energy $\epsilon$ and direction $\azimuthalAngle$ in $\vec{k}$-space.
Namely, the distributions $\fDiagonalE\left(\azimuthalAngle,\,\epsilon\right)$
and $\fSpinE\left(\azimuthalAngle,\,\epsilon\right)$ are derived
from the distributions $\fDiagonal\left(\vec{k}\right)$ and $\fSpin\left(\vec{k}\right)$,
and can again be writen as matrix $\fMatrixE\left(\azimuthalAngle,\,\epsilon\right)$.
Note that $\fMatrixE$ contains the spin-dependent DOS.  We now consider
a two-dimensional system, $\vec{k}=(k\cos\azimuthalAngle,\: k\sin\azimuthalAngle)$,
and assume that the spectrum $\energyOfK{k}$ in the absence of spin-orbit
interaction is isotropic {[}cf.\ Equation~(\ref{eq:Heffective}){]},
and we define $k_{\epsilon}$ such that $\energyOfK{k_{\epsilon}}=\epsilon$
and define $v_{\epsilon}=\energyOfK{k_{\epsilon}}'/\hbar$. For $\vec{E}=E\hat{\vec{x}}$
and for $b\ll E_{\mathrm{F}}$, the kinetic equation  becomes \cite{ShytovKinetic}
\begin{equation}
\frac{\partial\fMatrixE}{\partial t}+\bsigma\cdot\left[\frac{\vec{b}\times\fSpinE}{\hbar}-\frac{\fDiagonalE}{4\hbar^{2}v_{\epsilon}}\,\vec{b}\times\frac{\partial\vec{b}}{\partial k}\right]+\frac{eE}{\left(2\pi\right)^{2}}\,\frac{\partial f_{0}}{\partial\epsilon}\left[\frac{k_{x}}{\hbar}+\frac{1}{2\hbar^{2}\velocityE}\frac{\partial}{\partial\azimuthalAngle}\left(\vec{b}\cdot\bsigma\,\sin\azimuthalAngle\right)\right]={\left(\frac{\partial\fMatrixE}{\partial t}\right)\negthickspace,\negmedspace\negthickspace}_{\mathrm{coll.}}\label{eq:BlhsIntrinsic}\end{equation}
where $f_{0}$ is the Fermi distribution function and $\vec{b}$ is
evaluated for $\left|\vec{k}\right|=k_{\epsilon}$.

Now, considering only $\Hintrinsic$, the collision integral can be
found in a Golden Rule approximation \cite{ShytovKinetic},\begin{align}
\left(\frac{\partial\fMatrixE(\azimuthalAngle,\,\epsilon)}{\partial t}\right)_{\mathrm{coll}}= & \int_{0}^{2\pi}d\azimuthalAngle'\: K(\vartheta)\left[\fMatrixE(\azimuthalAngle')-\fMatrixE(\azimuthalAngle)\right]\nonumber \\
 & +\int_{0}^{2\pi}d\azimuthalAngle'\:\bsigma\cdot\left[\vec{M}\left(\azimuthalAngle,\azimuthalAngle'\right)\fDiagonalE\left(\azimuthalAngle'\right)-\vec{M}\left(\azimuthalAngle',\azimuthalAngle\right)\fDiagonalE\left(\azimuthalAngle\right)\right].\label{eq:CollIntrinsic}\end{align}
Here, the first term describes the spin-independent scattering, with
 $K\left(\vartheta\right)=K\left(\azimuthalAngle'-\azimuthalAngle\right)=W\left(q\right)k_{\epsilon}/2\pi\hbar^{2}v_{\epsilon}$
and $q=2k_{\epsilon}\sin\left(\left|\vartheta\right|/2\right)$. The
factor $W(q)=\big\langle\left|\Vimpurity(\vec{q})\right|^{2}\big\rangle$
does not depend on the direction of the momentum transfer $\vec{q}$
because the problem is isotropic (while of course the individual scattering
event is anisotropic, i.e., depends on the scattering angle $\vartheta$).
This spin-independent term coincides with first term of Equation~(\ref{eq:CrossSectionExtrinsic}),
with $K\left(\vartheta\right)=n_{i}v_{\epsilon}I\left(\vartheta\right)$,
and should only be included once. The second term in Equation~(\ref{eq:CollIntrinsic})
is given in first order in the intrinsic spin-orbit interaction $\vec{b}$
and contains the kernel \cite{ShytovKinetic}\begin{align}
\vec{M}\left(\azimuthalAngle,\azimuthalAngle'\right) & =\frac{v_{\epsilon}}{4k_{\epsilon}}\, K\left(\vartheta\right)\frac{\partial}{\partial\epsilon}\left[\frac{k_{\epsilon}\vec{b}\left(\azimuthalAngle\right)}{v_{\epsilon}}\right]+\frac{\vec{b}\left(\azimuthalAngle\right)+\vec{b}\left(\azimuthalAngle'\right)}{4\hbar k_{\epsilon}v_{\epsilon}}\,\frac{\partial K\left(\vartheta\right)}{\partial\vartheta}\,\tan\frac{\vartheta}{2},\end{align}
where the first term results from the spin-dependent DOS of the outgoing
wave.  The second term in $\vec{M}$ arises, because for a given
energy $\epsilon$, $\left|\vec{k}\right|$ depends on the spin state.
Thus the incoming and outgoing states can have different momenta,
leading to spin-dependent corrections to $q$. 

For a very smooth scattering potential such that typically $q<b/v_{\mathrm{F}}$,
the spin motion is adiabatic and should be treated differently \cite{GovorovBoundary,Khaetskii_2DGeneralized}.

\subsection{Diffusion equation}

\label{sub:Diffusive-equations} 
\newcommand{\tauSpinRelax}{\tau_{\mathrm{s}}}

\newcommand{\chargeDensityDiff}{N}

\newcommand{\spinDensity}{\vec{s}}

\newcommand{\spinDensityElem}{s}

\newcommand{\couplingSC}{\Gamma_{\mathrm{sc}}}

\newcommand{\couplingSS}{\Gamma_{\mathrm{ss}}}

\newcommand{\boundaryNormal}{\hat{\vec{n}}}

\newcommand{\Vdiff}{V_{E}}

A spin diffusion equation can be derived, starting from the Boltzmann
equation, for a dirty system when the spin relaxation time $\tauSpinRelax$
is much longer than the momentum relaxation time $\tau$, i.e., $\tauSpinRelax\gg\tau$.
It describes the carrier density $\chargeDensityDiff\left(\vec{r}\right)$
and spin polarization density $\spinDensity\left(\vec{r}\right)$;
say, $\spinDensityElem_{z}$ is the excess density of particles polarized
along $\hat{\vec{z}}$. For conduction band electrons, the (pseudo-)
spin density is $\vec{S}=\left(\hbar/2\right)\spinDensity$. The diffusion
equation is simpler to solve than the kinetic equation (\ref{eq:Kinetic}),
as the dependence on $\vec{k}$ is integrated out. Also, it is usually
sufficient to know $\spinDensity$, because it an experimentally accessible
quantity, while $\hat{f}(\vec{k})$ is not directly accessible, cf.\
Secs.~\ref{sub:Bulk-Spin-Polarization} and \ref{sub:Spin-at-boundaries}
below. For a two-dimensional system with Rashba spin-orbit interaction
$\HRashbaEl$ {[}Equation~(\ref{eq:RashbaEl}){]}, the diffusion
equation is \cite{Burkov04,Mish04} \begin{align}
\dot{\chargeDensityDiff} & =D\nabla^{2}\left(\chargeDensityDiff+\rho_{0}\Vdiff\right)+\couplingSC\left(\nabla\times\spinDensity\right)_{z},\label{eq:diffCarrier}\\
\dot{\spinDensityElem}_{i} & =D\nabla^{2}\spinDensityElem_{i}-\tau_{i}^{-1}\spinDensityElem_{i}+\couplingSS\left[\left(\hat{\vec{z}}\times\nabla\right)\times\spinDensity\right]_{i}+\couplingSC\:\left(\hat{\vec{z}}\times\nabla\right)_{i}\left(\chargeDensityDiff+\rho_{0}\Vdiff\right),\label{eq:diffSpin}\end{align}
with diffusion constant $D=\frac{1}{2}v_{\mathrm{F}}^{2}\tau$, anisotropic
Dyakonov-Perel \citeyear{DPrelax72} spin relaxation rates $\tau_{x}^{-1}=\tau_{y}^{-1}=\tau_{\perp}^{-1}=2\tau\left(\alpha k_{\mathrm{F}}/\hbar\right)^{2}$
and $\tau_{z}^{-1}=2\tau_{\perp}^{-1}$, spin-charge coupling $\couplingSC=-2\alpha\left(\alpha k_{\mathrm{F}}\tau\right)^{2}/\hbar^{3}$,
spin-spin coupling $\couplingSS=4\alpha E_{\mathrm{F}}\tau/\hbar^{2}$,
density of states $\rho_{0}=m/\pi\hbar^{2}$, and potential energy
$\Vdiff$ of a carrier in the electrical field. The charge current
is\begin{equation}
\vec{J}^{\mathrm{c}}=-D\nabla\left(\chargeDensityDiff+\rho_{0}\Vdiff\right)+\couplingSC\,\hat{\vec{z}}\times\spinDensity.\end{equation}
 Further, \citeasnoun{Malshukov_SpinAccumulation2DEGk3Dresselhaus}
derived diffusion equations for two-dimensional electrons with the
Dresselhaus Hamiltonian $\HDresselhausTwoDkkk$ {[}Equation~(\ref{eq:HDresselhaus2dk3}){]}.

The boundary conditions of the diffusion equation for a system with
spin-orbit interaction are not trivial and one expects that they depend
on the microscopic properties of the boundaries. A number of papers
have been written about the boundary conditions corresponding to various
physical circumstances and have partly clarified this issue \cite{GovorovBoundary,Malshukov_SpinAccumulation2DEGk3Dresselhaus,AdagideliBauer,GBS_boundary06,BleibaumBC,TserkovnyakBoundary}.
Depending on the particular boundary condition, there may or may not
be an $s_{z}$ spin accumulation near the boundary of a 2D system,
see Sec.~\ref{sub:Spin-at-boundaries}.  Somewhat related to these
question, \citeasnoun{Shekhter} considered the boundary between a
diffusive and ballistic system and allowed for spin-dependent scattering
at the boundary, resulting from a spatially dependent Rashba Hamiltonian
$\HRashbaEl$.

\section{Electrically induced spin polarization and spin transport}

\label{sec:Electrically-induced-SpinEffects}

\subsection{Spin current and spin Hall conductivity}

\label{sub:Spin-Current-and-Conductivity}
\newcommand{\spinCurrentSymbol}{\jSHc_{k}^{i}}
In this article, we define the spin current in a homogeneous system
with density $n$ as\begin{equation}
\spinCurrentSymbol\equiv\frac{1}{2}\, n\left\langle \sigma_{i}v_{k}+v_{k}\sigma_{i}\right\rangle ,\label{eq:SpinCurrent}\end{equation}
where $\left\langle \cdot\right\rangle $ is the expectation value
of single-particle operators and with $\left\langle 1\right\rangle =1$.
Thus, the spin current is defined as the difference of the particle
currents densities (measured in numbers of particles) for carriers
with opposite spins. This is in accordance with many studies \cite{MurakamiNZ03,Sinova04,Sinova_Workshop},
where a definition as in Equation (\ref{eq:SpinCurrent}) was chosen,
up to numerical prefactors. In many definitions, an additional prefactor
of $\frac{1}{2}$ is used, which results from $\hbar/2$ angular momentum
per electron spin and setting $\hbar=1$. With the same argument,
for the HH band, sometimes a prefactor $3/2$ is used; but sometimes
only a factor $\frac{1}{2}$ is used to have the same definition of
$\spinCurrentVec{\mu}$ for electrons and holes. Furthermore, the
r.h.s.\ of Equation (\ref{eq:SpinCurrent}) is sometimes multiplied
by the charge $e$ to obtain the same units for charge and for spin
currents. In particular, this means that the sign of the definition
of $\spinCurrentVec{\mu}$ may change if $e<0$ for electrons is taken.

Next, we define the spin Hall conductivity \begin{equation}
\spinHallConductivity\equiv-\jSHc_{y}^{z}/E_{x},\label{eq:SpinHallConductivity}\end{equation}
 where $\jSHc_{y}^{z}$ is the spin current density resulting from
a small applied electrical field $E_{x}$. The negative sign in Equation
(\ref{eq:SpinHallConductivity}) results from writing a formal similar
definition for $\spinHallConductivity$ as for the charge conductivity
$\sigma_{xy}$; however, sometimes a definition with an opposite sign
for $\spinHallConductivity$ is used. 

These various prefactors are only some technicality---the main question
is whether defining $\spinCurrentSymbol$ as in Equation (\ref{eq:SpinCurrent})
makes sense. To describe spin transport it sounds attractive to find
a scheme similar to the charge transport theory. Because of charge
conservation, charge densities $\rho^{\mathrm{c}}$ and charge currents
$\vec{j}^{\,\mathrm{c}}$ satisfy the continuity equation $\dot{\rho}^{\mathrm{c}}+\mathrm{div}\,\vec{j}^{\,\mathrm{c}}=0$.
For spin transport, we can consider the spin density $S_{i}$ instead
of $\rho^{\mathrm{c}}$. \possessivecite{Mott36} two-channel model
of electron transport in ferromagnetic metals is based on independent
and conserved currents of up- and down-spin electrons, and $\vec{S}$
and $\jSHc_{k}^{i}$ obey a continuity equation. The Definition (\ref{eq:SpinCurrent})
is the natural generalization of Mott's model; however, spin-orbit
coupling violates spin conservation, and the continuity equation for
spin densities and currents does not hold. In this article, we will
still use Equation (\ref{eq:SpinCurrent}) as definition of the spin
current, as it is widely used, but we remain aware of its limitations.
Despite the fact that it cannot be directly related to spin accumulation,
it is a useful model quantity to compare the effect of different spin-orbit
coupling mechanisms. While the continuity equation does not hold,
one can, for a concrete Hamiltonian, evaluate source terms arising
on the r.h.s.\  \cite{Burkov04,Erlingsson05}, which is often termed
as torque \cite{CulcerSpinCurrent}. 

Other definitions of spin currents have also been proposed. \citeasnoun{ZhangYang}
analyzed the current of the total angular momentum $L_{z}+S_{z}$
and argued that it vanishes for the Rashba Hamiltonian $\HRashbaEl$
in the absence of impurities (due to the rotational invariance of
$\HRashbaEl$) and that thus the impurity scattering would determine
angular momentum currents.  \citeasnoun{Zhang_SpinCurrentDef} discussed
spin currents, introducing a definition that is not proportional to
our $\spinCurrentSymbol$, but is given as time-derivative of the
{}``spin displacement'' $S_{i}\left(\vec{r}\right)\, r_{k}$. A
somewhat related definition was used by \citeasnoun{BryksinKleinert},
who found that such spin currents diverge when the frequency $\omega\to0$.

\subsection{Spin polarization}

\label{sub:Bulk-Spin-Polarization}

In experiments, the spin polarization can be detected optically \cite{OpticalOrientation}.
Electrically induced polarizations were inferred from measurements
of the Kerr rotation, where the polarization of a linearly polarized
beam of light rotates when the beam is reflected at a spin-polarized
sample \cite{KatoSpinHall,SihSHE}. Alternatively, the circular polarization
of the recombination light at a $p$-$n$ junction can be used to
determine the initial polarization of the carriers \cite{Wunderlich05}.
Finally, the inverse effect, the photo-galvanic effect, can be observed,
where a spin polarization is produced by polarized light and the induced
electrical current is detected \cite{GanichevPrettl}.

In the bulk of a two- or three-dimensional sample, spin polariziations
arise because an electrical field shifts the Fermi sea; $\left\langle \vec{k}\right\rangle =e\vec{E}\tau/\hbar$
for small $E$ and with transport lifetime $\tau$. This implies that
due to intrinsic terms, there is on average a finite spin-orbit field,
$\left\langle \vec{b}\left(\vec{k}\right)\right\rangle $. This leads
to a bulk spin polarization, which, in simple cases, is aligned is
along $\left\langle \vec{b}\right\rangle $ \cite{IvchenkoPikusSpinPol,VaskoPrima79,LNMSpinPolarization,Edelstein90,ALGPspinPol91}.
Such a polarization was observed experimentally in two-dimensional
GaAs hole systems: \citeasnoun{SilovBW04} used a $(001)$-surface
and detected the polarization of the photoluminescense from the side
of the cleaved sample, while \citeasnoun{GanichevPolarization} used
samples of low crystallographic symmetry and detected the polarization
along the growth direction. Furthermore, in the presence of anisotropic
scattering (see Sec.~\ref{sub:Kinetic-equations}), a magnetic field
$\vec{B}$ can lead to polarization perpendicular to both $\left\langle \vec{b}\right\rangle $
and $\vec{B}$ \cite{ERH_Polarzation}---such a perpendicular polarization
was already observed by \citeasnoun{KatoBulkPolarization04}.

\subsection{Spin currents in bulk}

\label{sub:Bulk-Spin-Currents}

The bulk spin current $\spinCurrentComponent{i}{k}$ was analyzed
for many different spin-orbit Hamiltonians. For a two-dimensional
electron system, the intrinsic effect of the Rashba coupling $\HRashbaEl$
lead to some debates. Because $\spinCurrentComponent{i}{k}$ is invariant
under time reversal, it is allowed to be finite in equilibrum. Indeed,
such equilibrum spin currents are predicted theoretically, however,
they are of order $\alpha^{3}$ and usually small \cite{R03}.

Now we discuss bulk spin currents driven by an external electrical
field for systems with either intrinisc or extrinsic spin-orbit interaction.
We do not discuss the more complicated case when both terms are present.

\begin{figure}
\begin{center}\includegraphics[%
  width=95mm,
  keepaspectratio]{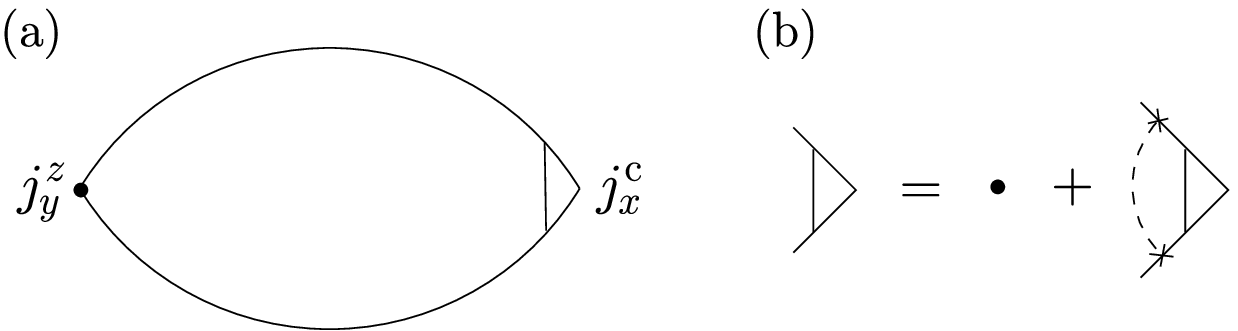}\end{center}

\caption{\label{cap:Diagram}Diagram for spin Hall conductivity. The Kubo
formula for $\spinHallConductivity$ is proportional to $\mathrm{Tr}\,\langle\!\langle\spinCurrentComponent{z}{y}G_{R}j_{x}^{\mathrm{c}}G_{A}\rangle\!\rangle$,
with charge current operator $j_{x}^{\mathrm{c}}$ and retarded and
advanced Green's functions $G_{R/A}$, and where $\langle\!\langle\cdot\rangle\!\rangle$
includes averaging over impurity configuration. In lowest order in
$1/k_{\mathrm{F}}\ell$, we can neglect crossed impurity lines and
$\spinHallConductivity$ is given by the diagram shown in (a). Here,
the full lines symbolize the renormalized Green's functions including
self-energy. In (b), the vertex renormalization due to impurity scatterings
(connected by dashed line) is defined recursively. When only the first
term of (b) is taken, we get the bubble contribution to $\spinHallConductivity$;
when all terms are summed, this leads to the additional (ladder) vertex
correction. }
\end{figure}

When an electrical field is applied and the electrons are accelerated,
the precession described in Sec.~\ref{sub:Intrinsic-Spin-Precession}
was considered. Because the initial spin density for a given direction
of $\vec{k}$ is proportional to $\alpha$, but the non-adiabatic
correction is proportional to $1/\alpha$, the spin-orbit coupling
constant cancels. Initially, it was believed that a small concentration
of impurities has no effect and a {}``universal'' spin Hall conductivity
$\spinHallConductivity=e/4\pi\hbar$ was predicted \cite{Sinova04}.
However, it turns out that when the impurities are properly taken
into account, the vertex correction cancels the bubble term, see Fig.~\ref{cap:Diagram}.
Thus, the dc conductivity $\spinHallConductivity$ vanishes  \cite{Inoue04,RaimondiSchwab_cancellationNumerics},
\begin{equation}
\spinHallConductivity=0,\label{eq:SHCZero}\end{equation}
which was confirmed in numerical calculations \cite{SSWH05}. Only
when the ac conductivity $\spinHallConductivity\left(\omega\right)$
is considered, in the regime $1/\tau\ll\omega\ll b/\hbar$ the universal
value is recovered \cite{Mish04}. 

 That there are no bulk spin Hall currents can be unterstood by
the following argument due to \citeasnoun{Dimitrova04}. Using the
Heisenberg equation and for parabolic bands, one finds the identity
$d\sigma_{y}/dt=-\left(m\alpha/\hbar^{2}\right)\left(\sigma_{z}v_{y}+v_{y}\sigma_{z}\right)$
for single particle operators \cite{Burkov04,Erlingsson05}. For a
homogeneous system, one then takes the expecation value of this identity
and finds that $\spinCurrentComponent{z}{y}\propto dS_{z}/dt$. When
we consider dc properties, we must assume that the system is in a
stationary state (i.e., we need impurity scattering). Then, the spin
polarization $S_{z}$ is constant and thus $\spinCurrentComponent{z}{y}=0$.
These arguments, as well as spin current cancellation in a magnetic
field in absence of scatterers \cite{RashbaSumRules04} show that
the cancellation is an intrinsic property the of free electron Hamiltonian
$\HRashbaEl$ and is not related to any specific property of the scatterers.
Furthermore, \citeasnoun{Chalaev2D} find that the weak localization
contribution to $\spinCurrentComponent{z}{y}$ vanishes, and show
more generally that $\spinCurrentComponent{z}{y}$ vanishes even if
both $\HRashbaEl$ and $\HDresselhausk$ are present. \citeasnoun{GrimaldiCM05}
find vanishing $\spinCurrentComponent{z}{y}$ for arbitrary values
of $\alpha k_{\mathrm{F}}/E_{\mathrm{F}}$.

This cancellation is special to $k$-linear spin-orbit interaction
(for nonparabolic bands, there can be a small finite contribution,
proportional to $\alpha^{2}$ \cite{KrotkovDasSarma_Nonparabolicity}).
For example, for two-dimensional hole systems, the coupling $\HRashbaHTwoD$
is cubic in $k$ {[}Equation (\ref{eq:HRashbaHTwoD}){]}. Then, if
isotropic scattering is assumed, the vertex correction vanishes, and
a {}``universal'' value $\spinHallConductivity=3e/4\pi\hbar$ is
found in the clean limit $b\tau\gg\hbar$ \cite{Murakami_NoHoleVertex,Schliemann_2DHolesk3}.
Quite generally, for a 2D system where the spin-orbit field $\vec{b}\left(\vec{k}\right)$
winds $N\neq1$ times around a circle in the $xy$ plane when $\vec{k}$
moves once around the Fermi circle (i.e., $N=3$ for $\HRashbaHTwoD$),
a universal value \begin{equation}
\spinHallConductivity=\frac{eN}{4\pi\hbar}\label{eq:SHCN}\end{equation}
 is found in the clean limit and for isotropic scattering \cite{ShytovKinetic,Khaetskii_2DGeneralized}.

Also for the $k^{3}$-Dresselhaus couplings, $\HDresselhausThreeD$
and $\HDresselhausTwoDkkk$ {[}Equations (\ref{eq:HDresselhausk})
and (\ref{eq:HDresselhaus2dk3}){]}, the vertex corrections vanish
for isotropic scattering. This leads to a finite spin Hall conductivity
for two dimensional systems when $\HDresselhausTwoDkkk$ is included
\cite{Malshukov_SpinCurrent2DEGk3Dresselhaus}. Similarly, $\spinHallConductivity$
is finite for $\HDresselhausThreeD$ \cite{BernevigZhang_3DDresselhaus}.

The results cited above are only valid for isotropic scattering, except
Equation (\ref{eq:SHCZero}), which holds more generally. Recently,
more general descriptions using kinetic equation (cf.\ Sec.~\ref{sub:Kinetic-equations})
allowed to include arbitrary angular dependence of impurity scattering
\cite{ShytovKinetic,Khaetskii_2DGeneralized}. It turns out that $\spinHallConductivity$
significantly depends on the shape of the scattering potential and
does not reduce to a simple form in general. For example, in the clean
limit $b\tau\gg\hbar$ and in the regime of small angle scattering
(but still for a typical momentum transfer $q>b/v_{\mathrm{F}}$,
i.e., not too small angles), one finds \begin{equation}
\spinHallConductivity=-\frac{eN}{2\pi\hbar}\left(\frac{N^{2}-1}{N^{2}+1}\right)\left(\tilde{N}-\zeta-2\right)\label{eq:SHCNsmallAngle}\end{equation}
 where $\zeta$ describes the non-parobalicity of the band, $v_{\epsilon}\propto k_{\epsilon}^{1+\zeta}$,
and for $\left|\vec{b}\right|\propto k^{\tilde{N}}$ \cite{ShytovKinetic}.
For example, taking $\zeta=0$ and $\HRashbaHTwoD$, with $N=\tilde{N}=3$
, we see that the sign of $\spinHallConductivity$ in Equation {[}Equation
(\ref{eq:SHCNsmallAngle}){]} is \emph{opposite} to the case of isotropic
scattering {[}Equation (\ref{eq:SHCN}){]}. Similarly, \citeasnoun{LiuLei}
found in a numerical study of system with spin-orbit coupling $\HRashbaHTwoD$
that $\spinHallConductivity$ strongly depends on the type of the
scattering potential. Therefore, the spin Hall conductivity is not
a universal quantity, as its numerical prefactor and its sign depend
on sample parameters. On the other hand, for clean systems with $N\neq1$,
the order of magnitude is consistently $\left|\spinHallConductivity\right|\sim e/4\pi\hbar$.

The above results are valid for weak spin orbit coupling, $k_{\alpha}\ll k_{\mathrm{F}}$.
Conversely, there are also materials with strong spin-orbit coupling,
as it was recently found in Bi/Ag (111) and Pb/Ag (111) surface alloys
\cite{AstBiAg}. This motivated \citeasnoun{GrimaldiCM05} to generalize
the theory; however, relying on an extensive numerical procedure.

The extrinsic contribution $\HextrinsicEl$ for electrons also leads
to spin currents \cite{DPpolarization,Hirsch99}. These currents 
are often evaluated only for isotropic impurity scattering \cite{Zhang00,ShchelushkinSHEmetals}.
Assuming absence of intrinisc spin-orbit interaction, for aribitrary
angular dependence of scattering, the extrinsic spin Hall conductivity
equals \cite{EHR_extrinsic} 

\begin{equation}
\spinHallConductivity=-\frac{\ShermanBmAvg}{2e}\,\sigma_{xx}+2n\SOcouplingEl\frac{e}{\hbar},\label{eq:SHCextrinsic}\end{equation}
where the first term is due to skew scattering (see Sec.~\ref{sub:Skew-Scattering}).
The second term due to the side-jump mechanism (Sec.~\ref{sub:Side-Jump-Mechanism});
as this mechanism goes beyond transport equation, this term has to
be evaluated separately. Here, $\sigma_{xx}$ is the electrical conductivity
and we defined the \emph{}\textit{\emph{transport skewness}} \textit{}

\begin{equation}
\ShermanBmAvg=\frac{\int d\Omega\: I\left(\vartheta\right)S\left(\vartheta\right)\sin\vartheta}{\int d\Omega\: I\left(\vartheta\right)\,\left(1-\cos\vartheta\right)}\label{eq:GammaSkew}\end{equation}
that depends on the structure of the scattering center and on the
Fermi energy, and $I$, $S$ are defined below in Equation~(\ref{eq:CrossSectionExtrinsic}).
For screened Coulomb scatterers, Equation~(\ref{eq:SHCextrinsic})
can be evaluated without any free parametrers \cite{EHR_extrinsic}
and the resulting absolute value of spin current is in quantitative
agreement (within error bars) with the observation by \citeasnoun{KatoSpinHall}
in GaAs and seems comparable with the data by \citeasnoun{SternZnSe}
in ZnSe---implying that the observed spin currents are due to the
extrinsic effect.  Note that in Equation~(\ref{eq:SHCextrinsic})
the skew scattering and the side jump contributions have opposite
signs. The skew scattering term dominates in standard transport theory
where one expands in $\hbar/E\tau$ where $E$ is a typical electron
energy; however, for dirty samples both terms can be of comparable
magnitude. \citeasnoun{SternZnSe} found that the measured $\spinHallConductivity$
in ZnSe has the sign of the skew scattering contribution and that
the same is likely to be the case for the $\spinHallConductivity$
observed by \citeasnoun{KatoSpinHall}. Finally, assuming short-range
scatterers, \citeasnoun{TseDasSarmaExtrinsic} found the same order
of magnitude for extrinsic spin currents. However; later they concluded
that intrinsic spin-orbit coupling can cancel skew-scattering and
reduce side-jump contributions to $\spinHallConductivity$ \cite{TseDasSarmaInEx}.

Remarkably, despite the fact that the side jump term in Equation~(\ref{eq:SHCextrinsic})
was derived by including electron dynamics during the scattering event,
it does not contain any factors related to the scattering probability---it
only depends on the coupling constant $\SOcouplingEl$, which is an
intrinsic property of the material and is directly related to a Berry
connection through the spin-orbit contribution to the operator of
electron coordinate, cf.\ Sec.~\ref{sub:Anomalous-velocity}. Thus,
while in this review and commonly in the literature the side-jump
contribution is considered as extrinsic, it is clear that the distinction
between intrinsic and extrinsic is somewhat arbitrary for this contribution.

\subsection{Anomalous Hall effect and its relation to spin Hall effect}

\label{sub:AHE}In the anomalous Hall effect (AHE), equilibrium polarization
of a ferromagnet combined with spin-orbit interaction leads to electrical
Hall currents transverse to an applied field.  The theory of AHE
has a long history and reveals many problems typical of spin transport
in media with spin-orbit coupling, including the competing mechanisms
of spin-orbit scattering by impurities and the role of intrinsic spin
precession; for reviews see \cite{Nozieres73,CrepieuxBrunoAHE,NagaosaAHE06}.
For non-interacting electrons and negligible spin relaxation, AHE
and SHE are closely related; this is true for extrinsic spin-orbit
interaction because $\SOcoupling$ is small and spin relaxation is
of order $\SOcoupling^{2}$ \cite{Elliott,YafetSSP}. In the SHE,
we can decompose the spin currents $\jSH^{\,\mu}$ as a difference
in particle currents of two spin species with polarizations $\pm\hat{\bvarepsilon}_{\mu}$.
Regarding these species separately, each carries the anomalous Hall
current $\JAHE^{\uparrow,\downarrow}$ of a system with spins fully
aligned along the $\pm\hat{\bvarepsilon}_{\mu}$ direction and with
density $n_{\mathrm{AH}}=\frac{1}{2}n$, because we consider the SHE
in non-magnetic media, where electrons are unpolarized in equilibrium.
We can express the spin Hall current as \cite{EHR_extrinsic}\begin{equation}
\jSH_{\,\mathrm{SH}}^{\,\mu}=e^{-1}\left(\JAHE^{\uparrow}-\JAHE^{\downarrow}\right).\label{eq:SHAH}\end{equation}
This relation allows to make use of the extensive literature on the
AHE to gain further insights into mechanisms of the SHE, at least
on its extrinsic part.

\subsection{Spin accumulation and transport at boundaries}

\label{sub:Spin-at-boundaries} For only extrinsic spin-orbit coupling,
because the spin relaxation is negligible for small $\SOcoupling$,
the spin is almost a conserved quantity and thus spin density and
spin current satisfy a continuity equation with a small relaxation
term. In this case, bulk spin currents will produce a spin polarization
at the edge of the sample, i.e., spin currents and spin accumulation
are directly related \cite{DPpolarization,Hirsch99,Zhang00}. The
polarization at a $y=0$ edge is $S_{z}=\left(\hbar/2\right)\spinDensityElem_{z}=\left(\hbar/2\right)\sqrt{\tau_{\mathrm{s}}/D_{\mathrm{s}}}\:\spinCurrentComponent{z}{y}$,
with spin relaxation time $\tau_{\mathrm{s}}$ and spin-diffusion
coefficient $D_{\mathrm{s}}$ (which is identical to the electron
diffusion coefficien $D$ in the absence of electron-elecon interaction).

For intrinsic spin-orbit interaction, it is not clear whether any
general relation exists between spin accumulation and bulk spin currents,
but spin accumulation can be studied directly. We discuss here the
situation of a semi-infinite two-dimensional electron system, with
Rashba coupling, located in the upper half-plane $(y>0)$. We assume
a uniform applied electric field parallel to the $x$-axis, and we
consider the diffusive limit, where the spin diffusion length is large
compared to the mean free path, so Eqs.~(\ref{eq:diffCarrier}) and
(\ref{eq:diffSpin}) apply far from the boundary. The spin density
near the edge will depend on the boundary conditions to the diffusion
equations at $y=0$, and these will depend, in turn, on the boundary
conditions of the microscopic Hamiltonian, as has been discussed in
the various articles cited in the last paragraph of Sec.~\ref{sub:Diffusive-equations}.

In the case of an ideal reflecting boundary, the spin density ${\bf {s}}(y)$
is found to be constant, and the same as in the bulk, right up to
the edge \cite{BleibaumBC}. Thus, one finds $s_{z}=0$, while $s_{y}$
has a value proportional to the charge current and to the Rashba coupling
constant $\alpha$. By contrast, if there is strong spin-orbit scattering
at the boundary, all components of ${\bf {s}}$ should vanish there.
In this case, the coupled diffusion equations predict that for $y>0$,
there will be \emph{non-zero} values of both $s_{z}$ and $s_{y}$,
with oscillating behavior, in a region near the edge whose width is
of the order of spin-precession length $k_{\alpha}^{-1}=\hbar^{2}/m\alpha$,
which is about the Dyakonov-Perel spin-diffusion length $\sqrt{D_{s}\tau_{s}}$
\cite{RashbaReview05}.

What happens if the boundary at $y=0$ is partially or completely
transmitting, and there is a second conductor in the region $y<0$
which has no spin orbit coupling? As noted by \citeasnoun{AdagideliBauer},
one should expect, in general, to find non-zero oscillatory values
of both $s_{z}$ and $s_{y}$ in the Rashba conductor near the boundary,
and injection of spin into the non-spin-orbit material. However, it
was found by \citeasnoun{TserkovnyakBoundary} that this actually
will not happen in the simplest case that one might consider: a boundary
between a pair of two-dimensional systems with identical properties
except for different values of $\alpha$. There will quite generally
be a discontinuity in the spin densities at a lateral boundary between
systems with different Rashba coupling. If the electron mobility is
a constant across the boundary, the discontinuity turns out to be
equal to the difference between the bulk spin-densities of the systems.
Then, on each side of the boundary one finds $s_{z}=0$, while $s_{y}$
is the same as the respective bulk value. There will thus be no spin
injection if the second two-dimensional electron system has $\alpha=0$.

\citeasnoun{Malshukov_SpinAccumulation2DEGk3Dresselhaus} have argued
that there should be spin accumulation near a reflecting edge, in
the diffusive case, when the cubic Dresselhaus coupling $\HDresselhausTwoDkkk$
is important. In the opposite limit of ballistic transport and near
a sharp specular edge, \citeasnoun{UsajB05} have found spin magnetization
due to $\HRashbaEl$ that oscillates rapidly with a period of $k_{\mathrm{F}}^{-1}$
and shows beating on a length scale of $k_{\alpha}^{-1}$. 

There are also numerical approaches analyzing the edge spin accumulation.
\citeasnoun{NomuraWunderlich} simulated a two-dimensional hole system,
using the coupling $\HRashbaHTwoD$ {[}Equation (\ref{eq:HRashbaHTwoD}){]}.
They found a spin accumulation that was consistent with the experiments
by \citeasnoun{Wunderlich05}.

\subsection{Mesoscopic systems and spin interferometers}

So far we have not considered interference effects---e.g., in Sec.~\ref{sub:Kinetic-equations},
we used an expansion in lowest order $1/k_{\mathrm{F}}\ell$ which
does not include interference between electron propagation paths that
follow different trajectories. To include such coherent effects in
systems with impurities, one needs to consider the next order in $1/k_{\mathrm{F}}\ell$:
the weak localization corrections. Alternatively, one can consider
clean systems that are ballistic on length scales of the device. Spin
interference effects in mesoscopic systems open up a new set of technical
possibilities, e.g., in rings or ring-like arrays with spin-orbit
interaction one can use Berry phase and Aharonov-Casher phase effects
to study a variety of phenomena. In the presence of an applied magnetic
field, spin-orbit effects can modify the Aharonov-Bohm or Altshuler-Aronov-Spivak
oscillations in the electrical conductance. For theoretical discussions,
see \cite{aronovLygeller,PhysRevLett_83_376,ELBerry,FrustagliaSpinFilter,KogaIFProposal,SigristRingRes}.
For experimental results, see \cite{morpurgo,ShayeganBerry,YangLGRing,KogaIF,KoenigAC,BagraevRing}.

For practical applications, it is not only important to generate nonequilibrium
spin polarization in media with spin orbit coupling, but also inject
spin currents produced by such populations into {}``normal'' conductors,
i.e., conductors with negligible spin orbit coupling. In normal conductors
spin is conserved and spin currents are well defined. Spin injection
can be achieved using spin-orbit coupling, even in devices without
magnetic fields and without ferromagnetic components. Proposals for
such devices, in the mesoscopic regime, have been made by \cite{JApplPhys_94_4001,Shekhter,SoumaNikolicRing,EtoSpinPol,SilvestrovMishchenko}.

Generally, spin interference devices make use of the intrinsic and
extrinsic spin-orbit couplings presented in Sec.~\ref{sec:Spin-Orbit-SC}
and the spin transport mechanisms discussed in Sec.~\ref{sec:SO-Mechanisms}
are important. However, we do not present more concrete descriptions
of interference effects or details of microscopic structures; this
could be done by solving the Schr\"odinger equation analytically,
by simulating it numerically, or by using weak localization calculations.
 On the other hand, we can assess the length scales on which spin
precession effects are expected: both diffusion equation (Sec.~\ref{sub:Diffusive-equations})
and its solution near boundaries (Sec.~\ref{sub:Spin-at-boundaries})
indicate spin precession length $\ell_{\alpha}=1/k_{\alpha}$ as a
characteristic length for spin distributions. Furthermore, for clean
systems, responses to an inhomogeneous field diverge at the wave vector
$q=2k_{\alpha}$ of the field \cite{RashbaResponse}. This {}``breakdown''
suggests the length $\ell_{\alpha}$ as an optimal size for achieving
large spin polarizations. For a more generic Hamiltonian $\Hintrinsic$,
this scale can be estimated as $\ell_{\textrm{eff}}\sim\hbar^{2}k_{F}/m\left|\vec{b}\right|$,
establishing a {}``mesoscopic'' scale at which one can expect largest
static spin responses to electric fields. Because $\ell_{\textrm{eff}}$
is also of the order of the Dyakonov-Perel spin diffusion length,
this estimate seems applicable both to the ballistic and diffusive
regimes.

\section{Spin Hall effect due to edge states in insulators}

\label{sec:SHEdgeStates} 
\newcommand{\Kpoint}{\mathrm{K}}

\newcommand{\Asite}{\mathrm{A}}

\newcommand{\Bsite}{\mathrm{B}}
 In the previous parts of Sec.~\ref{sec:Electrically-induced-SpinEffects},
we discussed spin currents $\spinCurrentComponent{z}{y}$ driven by
electric field $E_{x}$, their relevance to spin transport and spin
accumulation, and also the techniques for calculating conductivities
$\spinHallConductivity$ {[}Equation (\ref{eq:SpinHallConductivity}){]}.
Even when these nondiagonal components of the tensor $\spinCurrentComponent{i}{j}$
were not directly influenced by dissipation, they were calculated
for ordinary metallic conductors whose longitudinal electric conductivity
$\sigma_{xx}$ was controlled by electron scattering, hence, electron
transport in the bulk was dissipative. More recently, \citeasnoun{MurakamiNZ04}
proposed that some centrosymmetric 3D systems possess properties of
{}``spin insulators.'' These are media with gapped electron spectra
and zero bulk electrical conductivities $\sigma_{xx}$ but finite
and dissipationless spin conductivities $\spinHallConductivity$. 

The basic idea is as follows. A set of electron bands that in absence
of SO coupling belongs to orbital momentum $\vec{L}$, in presence
of SO coupling is described by the total angular momentum $\vec{J}=\vec{L}+\vec{S}$.
When some of the bands belonging to the $\vec{J}$ multiplet are filled,
while different bands of the same multiplet are empty and are separated
from the filled bands by a gap, all filled bands contribute to spin
current. Uniaxially strained zero-gap semiconductors $\alpha$-Sn
and HgTe, and narrow-gap semiconductors of PbTe type were proposed
as model systems. Spin conductivity $\spinHallConductivity$ is large
in these materials: it is about $e/\hbar a$ in 3D, $a$ being the
lattice constant. In 2D, $\spinHallConductivity$ is quantized when
the Fermi level is inside the gap \cite{QWZ05,ON05}. However, for
reasons similar to those discussed in Sec.~\ref{sub:Spin-Current-and-Conductivity},
the relation of this $\spinHallConductivity$ to spin transport is
not obvious and was already questioned \cite{KM05a,KM05b}. 

A different concept of spin transport in an insulating phase has been
developed by Kane and Mele as applied to graphene \cite{KM05a,KM05b}.
It is based on the Haldane model of quantum Hall effect (QHE) with
spinless fermions under the conditions of zero total magnetic flux
across the unit cell \cite{H88}. It has been emphasized \cite{ON05,KM05a}
that this model differers fundamentally from the model by \citeasnoun{MurakamiNZ04},
in particular in the properties of edge channels. Their crucial role
for the QHE has been clarified by \citeasnoun{H82}, and they play
a similar role in the physics of spin Hall effect in graphene. In
what follows, we consider properties of graphene in more detail. The
graphene model is not only of conceptual interest but is also attractive
because of the very recent experimental achievements in measuring
electron transport in graphene \cite{Geim04,Stormer05}. 

\begin{figure}
\begin{center}\includegraphics[%
  width=75mm,
  keepaspectratio]{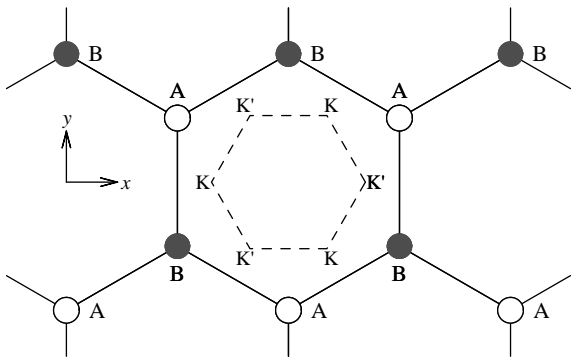}\end{center}

\caption{Schematics of honeycomb lattice of graphene. The hexagon in the center
is an elementary cell containing two carbon atoms that belong to two
sublattices. Atoms of these sublattices are marked as $\Asite$ and
$\Bsite$ and are shown by empty and filled circles, respectively.
Brillouin zone is shown by a dashed line. $\Kpoint$ and $\Kpoint^{\prime}$
indicate nonequivalent corners of the zone where the gap opens.}

\label{fig:honeycomb}
\end{figure}

Graphene is a monoatomic layer of graphite. Its honeycomb 2D lattice
is shown in Fig.~\ref{fig:honeycomb}. The elementary cell includes
two atoms shown as $\Asite$ and $\Bsite$. The phase diagram of graphene
can be understood from the tight-binding Hamiltonian \cite{KM05b}\begin{equation}
H=t\sum_{\langle ij\rangle}c_{i}^{+}c_{j}+\frac{2i}{\sqrt{3}}\lambda_{SO}\sum_{\langle\langle ij\rangle\rangle}\nu_{ij}c_{i}^{+}\sigma_{z}c_{j}+i\lambda_{R}\sum_{\langle ij\rangle}c_{i}^{+}(\bsigma\times\hat{\vec{d}}_{ij})_{z}c_{j}+\lambda_{v}\sum_{i}\xi_{i}c_{i}^{+}c_{i}\,,\label{eq3}\end{equation}
 where $c_{i}$ are annihilation operators at lattice cites $i$,
spin indeces in them being suppressed. The first term is the nearest
neighbor hopping term between two sublattices. For the following,
of the critical importance is the second term with $\nu_{ij}=(2/\sqrt{3})(\hat{\vec{d}}_{1}\times\hat{\vec{d}}_{2})_{z}=\pm1$
that describes second neighbor hopping. Here $\hat{\vec{d}}_{1}$
and $\hat{\vec{d}}_{2}$ are unit vectors along two bonds that an
electron traverses when going from the site $j$ to the site $i$.
The cross product of $\hat{\vec{d}}_{1}$ and $\hat{\vec{d}}_{2}$
produces a screw that in the Haldane model of spinless fermions couples
them to inhomogeneous magnetic flux, while in the present model it
couples the orbital motion of an electron to Pauli matrix $\sigma_{z}$.
Hence, $\lambda_{SO}$ is a coupling constant of a mirror symmetric,
$z\rightarrow-z$, spin-orbit interaction. The third term is a nearest
neighbor Rashba term, $\hat{\vec{d}}_{ij}$ being a unit vector in
the direction connecting $i$ and $j$ nodes. It explicitly violates
$z\rightarrow-z$ symmetry and originates from the coupling to the
substrate or from an external electric field. The fourth term is a
staggered sublattice potential with $\xi_{i}$ taking values $\xi_{i}=\pm1$
for $\Asite$ and $\Bsite$ lattice sites. It vanishes for graphene
but would be present for a similar boron nitride BN film. Including
this term is a clue for explaining the difference between the {}``quantum
spin Hall'' (QSH) phase and a usual insulator \cite{KM05a,KM05b}. 

\begin{figure}
\begin{center}\includegraphics[width=90mm,keepaspectratio]{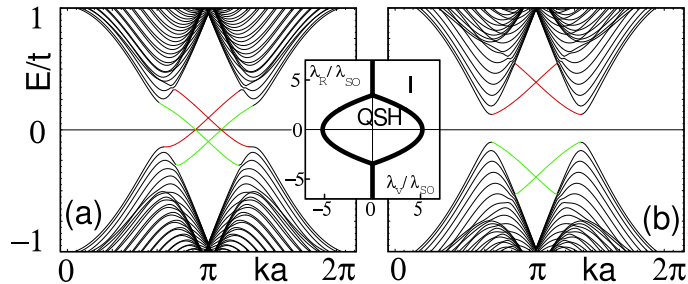}\end{center}

\caption[KM]{ \label{fig:KM} Energy bands of a one-dimensional strip
of graphene with {}``zig-zag'' edges, i.e., the strip has finite
extent along the $y$-direction shown in Fig.~\ref{fig:honeycomb}.
Narrow gaps in the bulk spectrum are achieved at the $\Kpoint$- and
$\Kpoint^{\prime}$-vertices of the Brillouin zone. Branches of the
spectrum originating from the edges of bulk continua show energies
of edge states. Edge states at a given edge of the strip cross at
$ka=\pi$, $a$ being the lattice spacing. (a) QSH phase for $\lambda_{v}=0.1t$;
edge states at a given edge of the strip cross at $ka=\pi$, $a$
being the lattice spacing. (b) A normal insulating phase for $\lambda_{v}=0.4t$.
In both cases $\lambda_{SO}=0.06t$ and $\lambda_{R}=0.05t$. The
inset shows the phase diagram in the $\lambda_{v}$-$\lambda_{R}$-plane
for $0<\lambda_{SO}\ll t$.  Reprinted figure with permission from [C.L. Kane and E.J. Mele, Phys.\ Rev.\ Lett.\ {\bf 95}, 146802 (2005)]. \htmladdnormallink{Copyright (2005) by the American Physical Society.}{http://link.aps.org/abstract/prl/v95/p146802}
}
\end{figure}

The remarkable spin properties of graphene are seen from the one-dimensional
projection of its energy spectrum, Fig.~\ref{fig:KM}, found by solving
the Hamiltonian of Eq.~(\ref{eq3}) in the geometry of a strip with
finite extent in the $y$-direction (defined in Fig.~\ref{fig:honeycomb}),
i.e., having {}``zig-zag'' edges aligned along the $x$-direction.
The spectrum comprises four energy bands, of which the two lower bands
are occupied; each bulk band is two-fold spin degenerate. Narrow gaps
at $\Kpoint$ and $\Kpoint^{\prime}$ open because of $\lambda_{SO},\lambda_{v}\neq0$---for
$\lambda_{SO},\lambda_{v}=0$, electrons possess a $k$-linear spectrum
of Dirac fermions, $\varepsilon(k)=\hbar c^{*}k$. In addition to
bulk states, there are edge states connecting $\Kpoint$ and $\Kpoint^{\prime}$
bulk continua. Their topology in the panels (a) and (b) is rather
different. 

In the panel (a) drawn for a small $\lambda_{v}$, edge states traverse
the gap. For each edge of the strip, there are two such states. They
are Kramers conjugate and propagate in opposite directions. This behavior
reflects unusual cross-symmetry of bulk states that manifests itself
in the opposite signs of the gap function at $\Kpoint$ and $\Kpoint^{\prime}$
points. The small-$\lambda_{v}$ phase has been dubbed as QSH-phase
by \citeasnoun{KM05a}. It is the distinctive property of this phase
that at any energy inside the gap there is one pair of edge modes
(more generally, an odd number of such pairs). 

When $\lambda_{R}=0$, $\sigma_{z}$ is conserved, and the pattern
of dissipationless spin transport become especially simple. Each of
independent subsystems of $\sigma_{z}=\uparrow$ and $\sigma_{z}=\downarrow$
electrons is equivalent to Haldane spinless fermions \cite{H88}.
One pair of such {}``spin filtered'' states propagates along each
edge. The states with opposite $\sigma_{z}$ polarizations propagate
in opposite directions. Because they form a Kramers doublet, potential
backscattering is forbidden and transport is dissipationless. This
model predicts two-terminal electric conductivity $G=2e^{2}/h$. Propagation
of charge current through edge states results in antisymmetric spin
accumulation at these edges. In four-terminal geometry, spin currents
flow between adjacent contacts, and related spin conductances are
quantized; when normalized on a number of transported spins, $G^{s}=\pm e/2\pi\hbar$.
For $\lambda_{R}\neq0$, however, $\sigma_{z}$ is not conserved.
Nevertheless, spin currents persist (if $\lambda_{R}$ is small, see
below) but they are no longer exactly quantized. An early argument
that the spin Hall conductance can be quantized was given by \citeasnoun{FroehlichLesHouches}
when considering incompressible 2D systems. 

Panel (b) of Fig.~\ref{fig:KM}, drawn for a larger staggered potential
$\lambda_{v}$, shows properties of a normal narrow gap insulator.
The gap function has the same sign at $\Kpoint$ and $\Kpoint^{\prime}$,
as a result, one pair of edge states runs between two conduction band
valleys and the second pair between two valence band valleys. For
some boundary conditions at the strip edges, edge states can penetrate
the gap. However, there is always an even number of Kramers conjugate
pairs of edge states at any given energy inside the gap, hence, backscattering
is no longer forbidden. Therefore, it is the topology of edge states
that defines the difference between the QSH and insulating phases
in a simple and concise form. 

The QSH phase is formed due to the bulk spin-orbit coupling $\lambda_{SO}$.
Increasing asymmetric $\lambda_{R}$ or staggered $\lambda_{v}$ potentials
destroy it when they become large enough. A phase diagram of the competing
phases, QSH phase and a normal insulator, is shown in the inset to
Fig.~\ref{fig:KM}. The QSH phase exists inside an ovaloid in the
$\lambda_{v}/\lambda_{SO}$--$\lambda_{R}/\lambda_{SO}$ plane. Outside
it, graphene shows properties of a normal narrow gap insulator. 

Another factor suppressing the gap and spin conductivity is electron
scattering in the bulk. Its effect has been investigated numerically
by Sheng et al. \cite{SSTH05} for a four-probe spin Hall setup by
using the Landauer-B\"{u}ttiker formula \cite{L88,B88_IBM}; their
spin conductivity $\spinHallConductivity$ describes real spin transport.
Disorder was modeled as $\sum_{i}\epsilon_{i}c_{i}^{+}c_{i}$ with
$\epsilon_{i}$ randomly distributed in the interval $[-W/2,W/2]$.
They found that $\spinHallConductivity$ remains within a few percent
of its quantized value when $W<t$ and the Fermi level stays inside
the gap; and  $\spinHallConductivity$ drops fast with increasing
$W$ for $W\gtrsim1.5t$. Under the same conditions, $\spinHallConductivity$
remains stable for $\lambda_{R}\lesssim0.2t$. These results allow
to establish the parameter range inside which edge spin channels remain
robust and carry dissipationless and nearly quantized spin currents.
Inside this range, there exist close analogy between QSH and QHE systems. 

Currently, there is no direct experimental indications of the spin
gap in graphene. A crude theoretical estimate of it by \citeasnoun{KM05a}
results in the gap $2\Delta_{\mathrm{SO}}\sim2.4$ K what is in a
reasonable agreement with different data \cite{BCP88}. However, more
recent calculations indicate that the actual value of $\Delta_{\mathrm{SO}}$
is considerably smaller. Meantime, estimates of $\hbar/\tau$ based
on transport data result in $\hbar/\tau\gtrsim25$ K for typical mobilities
of $\mu\approx10,000$ cm$^{2}$ V$^{-1}$ s$^{-1}$, $\tau$ being
the momentum relaxation time. Comparison of these data can easily
explain suppression of spin-polarized transport through edge channels
by disorder in the samples that are currently available. An independent
mechanism of suppression is the $\lambda_{R}$ constant that develops
when electron concentration is controlled by a gate, and the ratio
$\lambda_{R}/\lambda_{SO}$ is unknown. Unfortunately, all estimates
are crude because electron transport in graphene is still not understood.
\citeasnoun{NovDiracGraphene} recently reported the minimum metallic
conductivity $4e^{2}/h$ when the Fermi level passes through the conic
point of the spectrum; it is nearly independent of the mobility $\mu$.
A nonuniversal and even larger conductivity of about $6e^{2}/h$ was
reported by \cite{Stormer05}. The closest theoretical value $2e^{2}/h$
comes from the spin channel model \cite{KM05a}, followed by $4e^{2}/\pi h$
and $2e^{2}/\pi h$ found from different models of the bulk transport
of Dirac fermions \cite{Ziegler2DDirac,ShonAndo}. Because there exists
a region of parameter values where spin transport through edge channels
is robust, honeycomb lattices lithographically produced from semiconductors
with strong spin-orbit coupling may also be of interest \cite{ZhengAndo02}. 

\citeasnoun{AbaninLeeLevitov} argued that in the QHE regime, the
exchange-enhanced gap for chiral edge modes, originating from Zeeman
splitting, may be as large as $100\:\mathrm{K}$. Another system where
edge spin channels may play a role was proposed by \citeasnoun{BZ05};
it includes parabolic confinement in conjunction with inhomogeneous
shear deformation. More recently, \citeasnoun{FuKane} proposed $\mathrm{Bi}_{1-x}\mathrm{Sb}_{x}$
semiconductor alloys and $\alpha$-Sn and HgTe under uniaxial strain
as materials that satisfy the necessary symmetry requirements for
the QSH phase and that are expected to have a large $\Delta_{\mathrm{SO}}$.
Independently, \citeasnoun{BernevigHgTe} showed that HgTe/CdTe quantum
wells are also good candidates for the QSH phase.

\section{Acknowledgments}

We acknowledge discussions with A.A. Burkov,  D. Loss, A.H. MacDonald,
C.M. Marcus, E.G. Mishchenko, A.V. Shytov, and R. Winkler. This work
was supported by NSF Grants No. DMR-05-41988 and No. PHY-01-17795,
and the Harvard Center for Nanoscale Systems.

\newpage
\bibliographystyle{mamm}
\bibliography{S:/Physics/references}

\end{document}